\documentclass[letterpaper,12pt]{article}
\usepackage{amssymb,amsmath}
\usepackage{graphics}
\usepackage{graphicx}
\usepackage{braket}
\usepackage{tabularx} 
\usepackage{amsmath} 
\usepackage{graphicx}
\usepackage{setspace}
\usepackage[margin=1in,letterpaper]{geometry}
\usepackage{tikz}
\usepackage[italic]{hepnames}
\usepackage{float}
\usepackage{booktabs}
\usepackage{authblk}
\usepackage{color}
\usepackage[percent]{overpic}
\usepackage{subfig}
\usepackage{cite} 
\usepackage{appendix}
\usepackage[final]{hyperref} 
\hypersetup{
    colorlinks=true,      
    linkcolor=blue,       
    citecolor=red,        
    filecolor=magenta,     
    urlcolor=pink         
}
\begin{document}
\title{\textbf{Neutrinos as qubits and qutrits}}
\author{\normalsize{Abhishek Kumar Jha$^{1,2}$}\thanks{email: abhiecc.jha@gmail.com}}
\author{Akshay Chatla$^1$\thanks{email:chatlaakshay@gmail.com}}
\author{{Bindu A. Bambah$^1$}\thanks{email: bbambah@gmail.com}}
\affil{\small{$^1$School of Physics, University of Hyderabad, Hyderabad $500046$, India}}
\affil{\small{$^2$Department of Physics, Indian Institute of Science, Bengaluru $560012$, India}}
\date{\today}
\maketitle


\begin{abstract}
{We map neutrinos to qubit and qutrit states of quantum information theory by constructing the Poincar\'e sphere using SU(2) Pauli matrices and SU(3) Gell-Mann matrices, respectively. The construction of the Poincar\'e sphere in the two-qubit system enables us to construct the Bloch matrix, which yields valuable symmetries in the Bloch vector space of two neutrino systems. By identifying neutrinos with qutrits, we calculate the measures of qutrit entanglement for neutrinos. We use SU(3) Gell-Mann matrices tensor products to construct the Poincar\'e sphere of two qutrits neutrino systems. The comparison between the entanglement measures of bipartite qubits and bipartite qutrits in the two neutrino system are shown. The result warrants a study of two qutrits entanglement in the three neutrino system.\\ \\}

\end{abstract}

\section{Introduction}\label{sec1}

\label{Sec.1}
Quantum entanglement lies at the root of quantum information processing and quantum computation \cite{Michael:2020}. Quantum entanglement results from the non-classical correlations between separated quantum systems. Correlations between subsystems of a more extensive system that are not expressable in terms of correlation between local classical properties of the subsystem characterize quantum entanglement \cite{Caves:2000}. Linear independent quantum states can give rise to coherence and entanglement. A quantum system can be in any possible linear combination of multiple orthogonal states. A superposition of only two orthogonal states is called a qubit. It is a physical system described by a vectors of a two-dimensional Hilbert space $\mathcal{H}^2$. However, most quantum entanglement measures are realized between pairs of two qubits which lies in a 4-dimensional Hilbert space $\mathcal{H}^2\otimes\mathcal{H}^2$. Such measures include the partial transpose condition, which determines whether a state of two qubits is entangled, and other bipartite measures of entanglement such as concurrence and the entanglement of formation \cite{Horodecki:1997vt,Wootters:1997id,Coffman:1999jd}. Bipartite entanglement is limited in its applicability. Studying entangled states involving more than two qubits or multipartite entanglement opens up new possibilities in developing quantum theory and new quantum communication protocols. The simplest case of multipartite entanglement is tripartite entanglement in the three-qubit system. Examples of entangled tripartite states are the W states and the GHZ states of quantum optics. Genuine tripartite measures of entanglement such as the three-tangle and the three-$\pi$ have been used to characterize these states \cite{Yong:2007}.

Along with this type of tripartite entanglement, there has been an interest in generalizing the concept of a qubit to a qutrit. A qutrit is the superposition of three orthogonal states rather than the two which characterize a qubit. An operator representation of the qutrit density matrix has been developed, and qutrit entanglement has been studied in ref. \cite{Caves:2000}. Physically implementing a qutrit quantum computer in the context of trapped ions has been studied \cite{qtrit2} and quantum computer simulation packages for qutrits have been implemented \cite{qtrit3,qtrit4,qtrit5}. Furthermore, the generalized concurrence formula as a measure of two qutrits entanglement has also been studied \cite{Luthra}. 
 
Most of the systems in which quantum entanglement is studied are photonic or atomic systems. Recently, the ideas of quantum entanglement have been extended to the realm of particle physics by the study of two and three flavour neutrino systems {\cite{ Jha:2020hyh,Blasone:2007vw,Blasone:2007wp,Alok:2014gya,Song:2018bma,Naikoo:2017fos,Banerjee:2015mha}. Quantities such as the Leggett-Garg inequalities, whose spatial part is well known as Bell's inequality, can capture the violation of Bell-type inequality in the context of two and three flavour neutrino oscillations \cite{Formaggio:2016cuh,Naikoo:2019eec, Shafaq:2020sqo}. In {\cite{Jha:2020hyh}}, we place three-mode entanglement in neutrino oscillations on the same footing as mode entanglement in optical systems by mapping the neutrinos to the three mode W-state. We have studied tripartite measures such as the three-tangle and the three-$\pi$ for three flavour neutrino entanglement. Further study in this direction has shown that we are able to comprehend and forecast the behavior of neutrino oscillations by employing quantum information, and notably entanglement measures \cite{Quinta:2022sgq}.
 
 Two and three flavour neutrinos systems have been mapped to qubit states used in quantum information theory and encoded on an IBMQ computer using quantum computing as a tool \cite{Arguelles:2019phs,Molewski:2021}. It is generally known that the vacuum oscillations, the interaction with the surrounding matter, and the collective oscillations caused by interactions between various neutrinos all play a role in how the neutrino flavor changes in extreme astrophysical conditions. On a digital quantum computer of the most recent generation, the time-dependent many-body evolution of these astrophysical neutrinos has been investigated and showed the fundamental significance of error mitigation methods to derive meaning from entanglement measures utilising noisy, near-term quantum devices \cite{Hall:2021rbv}. In addition to this, the simulation of bipartite entanglement measures in the two flavour neutrino oscillation is shown on IBMQ processors \cite{Jha:2021}. Thus, one can say that quantum computing has been a good tool for studying neutrino oscillations and their entanglement properties.

Since the Hilbert space basis for three-flavour neutrino states has three dimensions, it cannot be fully represented using just one qubit. One possible approach, as suggested in ref.\cite{Arguelles:2019phs}, is to encode the minimal representation using two qubits. However, this leads to a redundant basis state that is not physically meaningful in the context of oscillating neutrinos. In order to eliminate this redundancy and achieve a more realistic representation, it is natural to consider a qutrit representation for neutrino flavours. This not only reduces the redundancy but also allows for a greater number of quantum operations to be performed in the neutrino system.

In this paper, we have presented the mathematical framework for neutrinos to be used in ternary system. We illustrate tripartite entanglement in neutrinos by considering them as ``qutrits". A qutrit is a linear superposition of three orthonormal basis states, $\ket{1}$, $\ket{2}$ and $\ket{3}$ : 
$\ket{\psi}=\alpha\ket{1}+\beta \ket{2}+\gamma\ket{3}$,
where $|{\alpha}|^2+|{\beta}|^2+|{\gamma}|^2=1$ and $\alpha,\beta,\gamma \in \mathbb{C}$. Since neutrino flavour states are a superposition of three states, it is only natural to try and characterize them as qutrits. We do this by mapping the density matrix for neutrinos to a generalized Poincar\'e sphere \cite{Caves:2000,Omar:2016}. In the two-flavour neutrino oscillation, geometric and topological phases such as the Berry's and Panchratnam's phase in terms of oscillation probabilities have been calculated using the Poincar\'e sphere and are directly observable\cite{Mehta:2009ea,Blasone:1999tq}. The Poincar\'e sphere has its origin in optics and is a way of visualizing different types of polarized light using the mapping from SU(2) to $S^3$. A qubit represents a point on the Poincar\'e sphere of SU(2) defined as, complex projective line $\mathcal{H}^2 = \mathcal{C}P^1 = SU(2)/U(1)$. A generalization of the Poincar\'e sphere to SU(3) can be constructed \cite{Arvind:1996rj,Mallesh:1997,Khanna:1997}. This construction has been the basis for characterizing qutrits that live in a 3-dimensional Hilbert space $\mathcal{H}^3$. A qutrit is represented by a point on the complex projective plane $\mathcal{H}^3 = \mathcal{C}P^2 = SU(3)/U(2)$ \cite{Bolukbasi:2006}. This work describes the entanglement of neutrinos by constructing Poincar\'e sphere representation for two- and three-flavour neutrino states using SU(2) Pauli matrices and SU(3) Gell-Mann matrices, respectively, to map the neutrino states to the qutrits states of quantum information theory.

The paper's organization is as follows: In Sect.(\ref{Sec.2}), we represent a single qubit density matrix of two flavour neutrino states in the basis of SU(2) Pauli matrices. In Sect.(\ref{Sec.3}), we use the tensor product of Pauli matrices as a basis to represent a two-qubit density matrix that led to the Bloch matrix construction in the two-qubit neutrino systems. We also show a bipartite entanglement measure concurrence quantification in the two neutrino system. In Sect.(\ref{Sec.4}), we describe a qutrit density matrix of three flavour neutrino states in the basis of SU(3) Gell-Mann matrices. Under a particular set of constraints, the measure for entanglement characterized by the entropy of mixing for a three flavour neutrino system is found.
In Sect.(\ref{Sec.5}), we represent a two qutrit density matrix in the basis of SU(3) Gell-Mann matrices tensor products which describe the generalized matrix construction for two qutrits neutrino states. Furthermore, we compute generalized concurrence as a measure of bipartite qutrits entanglement in the two neutrino system and compare it with the concurrence of the bipartite qubit neutrino system. Finally, the conclusion is given in Sect.(\ref{Sec.6}).  

\section{SU(2) Poincar\'e sphere for two-flavour neutrinos}
\label{Sec.2}
                                                                                                                                                                                                                                                                                                                                                                                                                                                                                                                                                                                                                                                                                                                                                                                                                                                                                                                                                                                                                                                                                                                                                                                                                                                                                                                                                                                                                                                                                                                                                                                                                                                                                                                                                                                                                                                                                                                                                                                                                                                                                                                                                                                                                                                                                                                                                                                    
 In \cite{ Jha:2020hyh}, the two-flavour neutrino system has been mapped onto a qubit system in quantum optics, with the two flavours mixing matrix playing the role of the beam splitter in a two-level quantum optical system. It is well known that a two-level system can be mapped to the Poincar\'e sphere of two
level quantum systems \cite{Arvind:1996rj}. 
For a two-dimensional complex Hilbert space $\mathcal{H}^2$, a quantum state $\ket{\psi}$ can be written as a superposition 
\begin{equation}
\ket{\psi}=c_1 \ket{0}+ c_2\ket{1})
\end{equation}
where $|c_1|^2+|c_2|^2=1$ and $ c_1, c_2 \in\mathbb{C}$ 
\begin{eqnarray}
        \ket{0}=\begin{pmatrix}
        1\\
        0\\
        \end{pmatrix}; \ket{1}=\begin{pmatrix}
        0\\
        1\\
        \end{pmatrix}.
    \end{eqnarray}
    Using the polar representation $c_1=r_1e^{i\varphi_0}$ and $c_2=r_2e^{i\varphi_1}$ and the fact that, in the case of quantum bits, a quantum state $\ket{\psi}$ does not change if multiplied by an overall phase $e^{-i\varphi_0}$ the equivalent quantum state is 
 \begin{equation}
     e^{-i\varphi_0} \ket{\psi}=r_1 \ket{0}+ r_2e^{i\varphi_1-i\varphi_0}\ket{1}.
 \end{equation}    
 Using the angular representation of complex variables and the fact that $r_1^2+r_2^2=1$ and $\phi= \varphi_1-\varphi_0 $ we get a representation of the equivalent representation of $\ket{\psi}$ as 
\begin{equation}
\ket{\psi}={\cos}(\theta) \ket{0}+ {\sin}(\theta)e^{i\varphi}\ket{1}.
\end{equation}
For a two-dimensional complex Hilbert space $\mathcal{H}^2$, the density matrix correspond to a pure state $\ket{\psi}$ is given by $\rho=\ket{\psi}\bra{\psi}$. Its expansion in terms of Pauli matrices $\sigma_j$ leads to the Poincar\'e sphere construction 
    \begin{equation}
        \rho=\ket{\psi}\bra{\psi}=\frac{1}{2}(1+\hat{n}.\Vec{\sigma}),
        \label{Eq.5}
    \end{equation}
    where $\rho^{\dagger}=\rho^2=\rho \ge 0$, $Tr\rho=1$ $\Longrightarrow$ $\hat{n}^*=\hat{n}$, $\hat{n}.\hat{n}=1$ $\Longleftrightarrow$ $\hat{n}\in S^2$ is the unit vector on the sphere. Thus there is a one to one correspondence between pure qubit states and points on the unit sphere $S^2$ embedded in $\mathcal{R}^3$, which is known as the Poincar\'e sphere construction (of which the Bloch sphere is a particular case).  
  If  $\ket{\psi^{\prime}}$ and $\ket{\psi}$ are two pure states then 
    \begin{equation}
     Tr(\rho^{\prime}\rho)=|<{\psi^{\prime}}|{\psi}>|^2=\frac{1}{2}(1+{\hat{n}}^{\prime}.\hat{n}),
      \end{equation}
      where ${\hat{n}}^{\prime}$ is the unit vector on the sphere corresponding to $\ket{\psi^{\prime}}$.  For orthogonal states $|<{\psi^{\prime}}|{\psi}>|^2=0$, so that $ 1+{\hat{n}}^{\prime}.\hat{n}=0$ and thus correspond to the diametrically opposite point on $S^2$ correspond to mutually orthogonal Hilbert space vectors where $<\psi^{\prime}|\psi>$ is the inner product in $\mathcal{H}^2$. 
    Applying an SU(2) transformation to $\ket{\psi}\in \mathcal{H}^2$ the representative point in SU(2) $\hat{n}\in S^2$ (circle) undergoes a rotation belonging to $SO(3)$
    \begin{eqnarray}
        &{\ket{\psi^{'}}=u\ket{\psi}, u\in SU(2) \Longrightarrow  n_j^{\prime} =R_{jk}(u)n_k;}&\nonumber\\
        &{ R_{jk}(u)=\frac{1}{2} Tr(\sigma_j u \sigma_k u^{\dagger}),}& 
     \end{eqnarray}
    $R(u)\in SO(3)$.
     Thus all elements $R\in SO(3)$ are realized in this way, and we have the coset space identifications (since multiplication by a phase leads to equivalent representations) $S^2=SU(2)/U(1)=SO(3)/SO(2)$ \cite{Arvind:1996rj}.
     
     Two-flavour neutrino oscillations involve a Hilbert space of two dimensions $\mathcal{H}^2$, and the mixing matrix is given by the SU(2) matrix \cite{Carlo:2010}. Let the mass eigenstates of the two flavour neutrino system be $\ket{\nu_1}$ and $\ket{\nu_2}$ then using the mixing matrix $U(\theta)$ (where $U^{*}(\theta)$ is complex conjugate of $U(\theta)$), the flavour state $\ket{\nu_e}$ and $\ket{\nu_\mu}$ can be written in linear superposition of mass eigenstates ($\ket{\nu_1}$, $\ket{\nu_2}$) basis  as
     
      \begin{eqnarray}
&{\begin{pmatrix}
    \ket{\nu_e}\\
    \ket{\nu_\mu}\\
    \end{pmatrix}=U^*(\theta)\begin{pmatrix}
    \ket{\nu_1}\\
    \ket{\nu_2}\\
    \end{pmatrix};}&\nonumber\\
    &{\text{where,}\hspace{0.1cm} U(\theta)=\begin{pmatrix}
    {\cos}\theta & {\sin}\theta \\
    -{\sin}\theta & {\cos}\theta \\
    \end{pmatrix}\in SU(2).}&
    \label{Eq.8}
\end{eqnarray}
 Then, the time evolved flavour neutrino states in linear superposition of two mass eigenstate basis are 
    \begin{eqnarray}
        {\ket{\nu_e(t)}={\cos}\theta e^{-iE_1t/\hbar}\ket{\nu_1}-{\sin}\theta e^{-iE_2t/\hbar}\ket{\nu_2},}&\\ \nonumber
        {\ket{\nu_\mu(t)}={\sin}\theta e^{-iE_1t/\hbar}\ket{\nu_1}+{\cos}\theta e^{-iE_2t/\hbar}\ket{\nu_2}.}&
     \end{eqnarray}  
$\ket{\nu_e(t)}$ can be parametrized by two angles $\theta$ and $\phi$ as
            \begin{eqnarray}
               & {\ket{\nu_e (\theta,\phi)}=e^{-iE_1t/\hbar}({\cos}\theta\ket{\nu_1} -{\sin}\theta e^{-i(E_2-E_1)t/\hbar}\ket{\nu_2}),}&\nonumber\\
               & {=e^{-iE_1t/\hbar}({\cos}\theta \ket{\nu_1}-{\sin}\theta e^{-i\phi}\ket{\nu_2}),}&
            \end{eqnarray}
            where $E_1=(p^2+m_1^2)^{1/2}$, $E_2=(p^2+m_2^2)^{1/2}$ and  $\phi=\frac{(E_2-E_1)t}{\hbar}=\frac{\Delta m^2 t}{2E\hbar}$, $\Delta m^2\equiv m^2_2-m^2_1$. The overall phase is redundant and leads to an equivalent representation in such a way that the coefficient of $\ket{\nu_1}$ is real. Thus, the normalized time evolved electron neutrino and muon neutrino flavour state are,
            \begin{eqnarray}
                 {\ket{\nu_e (\theta,\phi)}={\cos}\theta \ket{\nu_1}-{\sin}\theta e^{-i\phi}\ket{\nu_2},}&\nonumber\\
                { \ket{\nu_{\mu} (\theta,\phi)}={\sin}\theta \ket{\nu_1}+{\cos}\theta e^{-i\phi}\ket{\nu_2},}&
                \label{Eq.11}
            \end{eqnarray}
            respectively. Now we can easily identify the mass eigenstates of a flavour neutrino state to the qubit states 
    \begin{eqnarray}
        \ket{0}=\ket{\nu_1}=\begin{pmatrix}
        1\\
        0\\
        \end{pmatrix}; \ket{1}=\ket{\nu_2}=\begin{pmatrix}
        0\\
        1\\
        \end{pmatrix}.
    \end{eqnarray} 
    Identifying the states  $\ket{\psi}$ and $\ket{\psi^{\prime}}$ with time evolved flavour neutrino states  $\ket{\nu_e (\theta,\phi)}=\begin{pmatrix}
            {\cos}\theta\\
            -e^{-i\phi}{\sin}\theta\\
            \end{pmatrix}$ and, $ \ket{\nu_{\mu} (\theta,\phi)}=\begin{pmatrix}
            {\sin}\theta\\
            e^{-i\phi}{\cos}\theta\\
            \end{pmatrix}$, we see that  $\ket{\nu_e(\theta,\phi)}$ is an eigenstate with eigenvalue +1.
  
\begin{equation}
    \hat{O}=\hat{n}(\theta,\phi).\Vec{\sigma}=\begin{pmatrix} 
            {\cos}2\theta & -{\sin}2\theta e^{i\phi}\\
            -{\sin}2\theta e^{-i\phi} & -{\cos}2\theta\\
            \end{pmatrix}\in SU(2).
\end{equation}            
             
             Here $\Vec{\sigma}=(\sigma_1,\sigma_2,\sigma_3)$ and $\hat{n}(\theta,\phi)$ is a real unit vector, $\hat{n}(\theta,\phi)=-{\sin}2\theta {\cos}\phi \hat{e_1}+{\sin}2\theta {\sin}\phi\hat{e_2}+ {\cos}2\theta\hat{e_3}$ called the Poincar\'e unit vector. Therefore,
\begin{eqnarray}
           { \hat{O}\ket{\nu_e(\theta,\phi)}=\ket{\nu_e(\theta,\phi)}.}&
            \end{eqnarray} 
            Thus, a state $\ket{\nu_e(\theta,\phi)}\in \mathcal{H}^2$ is expressed in terms of a unit vector $\hat{n}(\theta,\phi)$ on the surface of the Poincar\'e sphere. This correspondence is one-to-one if the ranges of $\theta$ and $\phi$ are restricted to $0\leq\theta\leq\pi$ and $0\leq\phi<2\pi$. The $2\times 2$ density matrix is given by 
            \begin{equation}
                \rho^{e}_{2\times2}=\begin{pmatrix}
                {\cos}^2\theta & -e^{i\phi} {\sin}\theta cos\theta\\
               - e^{-i\phi}{\sin}\theta {\cos}\theta & {\sin}^2\theta\\
                \end{pmatrix}=\frac{1}{2}(I+\hat{n}.\Vec{\sigma}),
                \label{Eq.16}
            \end{equation}
          which is the same as Eq.(\ref{Eq.5}).
            The eigenvalues of $\rho^e_{2\times2}$ are 1 and 0, therefore $\rho^e_{2\times2}$ is a rank 1 density matrix. This maps the neutrino state $\ket{\nu_e(t)}$ to the the surface of the unit sphere in the three dimensional vector space. A similar mapping can be done for the neutrino state $\ket{\nu_{\mu}(t)}$. The density matrix correspond to $\ket{\nu_{\mu}(\theta,\phi)}$ is
 \begin{equation}
 \rho^{\mu}_{2\times2}=\begin{pmatrix}
 {\sin}^2\theta & e^{i\phi} {\sin}\theta {\cos}\theta\\
 e^{-i\phi} {\sin}\theta {\cos}\theta & {\cos}^2\theta\\
 \end{pmatrix}=\frac{1}{2}(I+\hat{n}^\prime.\Vec{\sigma}),
 \label{Eq.17}
 \end{equation}
 where $\hat{n}^\prime(\theta,\phi)={\sin}2\theta {\cos}\phi \hat{e_1}- {\sin}2\theta {\sin}\phi\hat{e_2}- {\cos}2\theta\hat{e_3}$. When $\theta\rightarrow \frac{\theta}{2}$ then the Poincar\'e sphere becomes the Bloch sphere used in quantum optics. In the next section, we use Eq.(\ref{Eq.16}) and Eq.(\ref{Eq.17}) to describe the Bloch vector and its generalized representation in the two-qubit neutrino systems.

 \section{Bloch matrix construction of two qubit neutrino states}
\label{Sec.3}
                                                                                                                                                                                                                                                                                                                                                                                                                                                                                                                                                                                                                                                                                                                                                                                                                                                                                                                                                                                                                                                                                                                                                                                                                                                                                                                                                                                                                                                                                                                                                                                                                                                                                                                                                                                                                                                                                                                                                                                                                                                                                                                                                                                                                                                                                                                                                                                    
A 4$\times$4 density matrix $\rho_{4\times 4}\in\mathcal{H}^2\otimes\mathcal{H}^2$ represent either a single four-level system, or a pair of coupled two-level systems \cite{Omar:2016}: two qubits. The study of the Bloch matrix using the density matrix $\rho_{4\times 4}$ will give useful symmetries in the Bloch-vector space. This section studies the two-qubit density matrices of two-flavour neutrino states in the Dirac-basis to construct the Bloch-matrix. We extend this idea to study the entanglement nature of two-qubit neutrino systems.

 In general, any $2\times2$ density matrix $\rho_{2\times2}$ of a single qubit state can be represented as in the Pauli basis as
\begin{equation}
\rho_{2\times2}=\frac{1}{2}(1+\vec{u}.\vec{\sigma})=\frac{1}{2}r_\mu\sigma_\mu,
\label{Eq.18}
\end{equation}
where the scalar coefficients $r_\mu=Tr(\rho_{2\times2}\sigma_\mu)$ ($\mu=0,1,2,3$) in which $r_0$ is always unity to ensure $Tr\rho_{2\times2}=1$, and $r_1$, $r_2$, $r_3$ are the components of the Bloch vector $\vec{u}$, and $\sigma_\mu$ are the Pauli matrices. Similarly, using Eq.(\ref{Eq.18}), the density matrix $\rho_{4\times4}$ of any two-qubit states can be constructed using the Dirac matrices, denoted $D_{\mu \nu}=\sigma_\mu\otimes\sigma_\nu$ (the tensor product of two Pauli matrices) as its basis such that 
    \begin{equation}
        \rho_{4\times4}=\frac{1}{4}r_{\mu \nu}D_{\mu \nu},
        \label{Eq.19}
    \end{equation}
    where $\mu,\nu=0,1,2,3$. The characterization of the Pauli matrices and Dirac matrices are shown in ref. \cite{Omar:2016}. The scalar coefficients $r_{\mu \nu}$ is defined as
        \begin{equation}
        r_{\mu \nu}=Tr(\rho_{4\times4}D_{\mu\nu})=<\sigma_\mu\otimes\sigma_\nu>
        \label{Eq.20}
        \end{equation}
        constitute 16 entries of the Bloch matrix $\textbf{M}$ \cite{Omar:2016}.
        The Bloch matrix $\textbf{M}$ is split into four components: a scalar of unity, two three-dimensional vectors, and a 3$\times$3 matrix. We write
         \begin{equation}
\textbf{M}=\left[\begin{array}{c|ccc}

1 & r_{01} & r_{02} & r_{03} \\

\hline

r_{10} & r_{11} & r_{12} & r_{13}\\
r_{20} & r_{21} & r_{22} & r_{23}\\
r_{30}& r_{31} & r_{32} & r_{33}\\
\end{array}\right ],
\label{Eq.21}
\end{equation}   
        where $u_i=r_{i0}$ and $v_j=r_{0j}$ ($i,j=1,2,3$) are the components of two local Bloch vectors $\vec{u}$ and $\vec{v}$, respectively. $R_{ij}=r_{ij}$ is the matrix elements of correlation matrix R, and $r_{00}=1$ implies $\rho_{4\times4}$ be a Hermitian matrix, of unit trace, and positive semidefinite. 
 
 Using Eq.(\ref{Eq.16}) and Eq.(\ref{Eq.17}), we construct the density matrix of two qubit neutrino states $\ket{\nu_{e\mu}(\theta,\phi)}=\ket{\nu_e(\theta,\phi)}\otimes\ket{\nu_\mu(\theta,\phi)}$ in the standard basis ($\ket{0}\otimes\ket{0}\equiv\ket{00},\ket{0}\otimes\ket{1}\equiv\ket{01},\ket{1}\otimes\ket{0}\equiv\ket{10},\ket{1}\otimes\ket{1}\equiv\ket{11}$) as

\begin{eqnarray}
&{\rho^{e\mu}_{4\times4}=\rho^{e}_{2\times2}\otimes\rho^{\mu}_{2\times2}=\ket{\nu_{e\mu}(\theta,\phi)}\bra{\nu_{e\mu}(\theta,\phi)}}& \label{Eq.22}\\
&{
=\begin{pmatrix}
{\cos}^2\theta {\sin}^2\theta & e^{i\phi}{\cos}^3\theta {\sin}\theta & -e^{i\phi}{\cos}\theta {\sin}^3\theta & -e^{2i\phi}{\cos}^2\theta {\sin}^2\theta\\
e^{-i\phi}{\cos}^3\theta {\sin}\theta & {\cos}^4\theta & -{\cos}^2\theta {\sin}^2\theta & -e^{i\phi}{\cos}^3\theta {\sin}\theta \\
-e^{-i\phi} {\cos}\theta {\sin}^3\theta & -{\sin}^2\theta {\cos}^2\theta & {\sin}^4\theta & e^{i\phi}{\cos}\theta {\sin}^3\theta\\
-e^{-2i\phi}{\sin}^2\theta {\cos}^2\theta & -e^{-i\phi} {\sin}\theta {\cos}^3\theta & e^{-i\phi}{\cos}\theta {\sin}^3\theta & {\sin}^2\theta {\cos}^2\theta \\
\end{pmatrix}}&\nonumber
\end{eqnarray}

 We can expand the above two qubit density matrix $\rho^{e\mu}_{4\times4}$ uniquely as 
\begin{eqnarray}
          & {\rho^{e\mu}_{4\times4}=\rho^e_{2\times2}\otimes\rho^\mu_{2\times2}=\frac{1}{4}[(I+\hat{n}.\Vec{\sigma}^e)\otimes(I+\hat{n}^\prime.\Vec{\sigma}^{\mu})]}&\nonumber\\
&{=\frac{1}{4}[I\otimes I+\vec{\sigma}^e .\hat{n}\otimes I+ I \otimes \vec{\sigma}^{\mu}.\hat{n}^\prime + \sum_{i,j=1}^3 r_{ij}\sigma^e_i\otimes\sigma^{\mu}_j],\hspace{0.7cm}}&
\label{23}
\end{eqnarray}
 the expansion coefficients are 
\begin{eqnarray}
{n_i=tr(\rho^{e\mu}_{4\times4}\sigma_i\otimes I),}&\nonumber\\
{n^\prime_j= tr(\rho^{e\mu}_{4\times4} I\otimes \sigma_j),}&\nonumber\\
{r_{ij}= tr(\rho^{e\mu}_{4\times4} \sigma_i\otimes\sigma_j),}&
\label{Eq.24}
\end{eqnarray}
where $i,j=1,2,3$.
In Eq.(\ref{Eq.24}), ${n}_i$ and  ${n}^\prime_j$ are the elements of Poincar\'e unit vector $\hat{n}$ and $\hat{n}^\prime$, respectively and the coefficients $r_{ij}$ of the basis $\sigma_i\otimes\sigma_j$ is defined as a correlation matrix R between the two sub-system $\rho^e$ and $\rho^\mu$ as
\begin{equation}
R=\begin{pmatrix}
r_{11}&r_{12}&r_{13}\\
r_{21}&r_{22}&r_{23}\\
r_{31}&r_{32}&r_{33}\\
\end{pmatrix}.
\label{Eq.25}
\end{equation}
Using Eq.(\ref{Eq.24}), the elements of $r_{ij}$ can be obtained as:  $r_{11}=-4 {\cos}^2\theta {\sin}^2\theta {\cos}^2\phi$, $r_{12}=r_{21}=2 {\cos}^2\theta {\sin}^2\theta {\sin}2\phi$, $r_{13}=r_{31}=\frac{1}{2}{\sin}4\theta {\cos}\phi$, $r_{22}=-{\sin}^22\theta {\sin}^2\phi$, $r_{23}=r_{32}=-\frac{1}{2}{\sin}4\theta {\sin}\phi$, $r_{33}=-{\cos}^22\theta$. Alternative representation of Eq.(\ref{23}) is Eq.(\ref{Eq.19}). So, we can incorporate this correlation matrix R (see Eq.(\ref{Eq.25})) into the Bloch-matrix \textbf{M} shown in Eq.(\ref{Eq.21}). 

        Following Eq.(\ref{Eq.19}) and using Eq.(\ref{Eq.22}) in Eq.(\ref{Eq.20}), the Bloch matrix for $\rho^{e\mu}_{4\times4}$ can be constructed as
        
\begin{equation}
\textbf{M}_{e\mu}=\left[\begin{array}{c|ccc}
1 & {\cos}\phi {\sin}2\theta & -{\sin}2\theta {\sin}\phi  & -{\cos}2\theta\\

\hline

        -{\cos}\phi {\sin}2\theta & -4{\cos}^2\theta {\cos}^2\phi {\sin}^2\theta & 2{\cos}^2\theta {\sin}^2\theta {\sin}2\phi & \frac{1}{2}{\cos}\phi {\sin}4\theta\\
        {\sin}2\theta {\sin}\phi & 2{\cos}^2\theta {\sin}^2\theta {\sin}2\phi & -{\sin}^2 2\theta {\sin}^2\phi & -\frac{1}{2}{\sin}4\theta {\sin}\phi\\
        {\cos}2\theta & \frac{1}{2}{\cos}\phi {\sin}4\theta & -\frac{1}{2}{\sin}4\theta {\sin}\phi & -{\cos}^2 2\theta\\
\end{array}\right ].
\label{Eq.26}
\end{equation}        
        
 By comparing the matrix elements of $\textbf{M}_{e\mu}$ (see Eq.(\ref{Eq.26})) with $\textbf{M}$ (see Eq.(\ref{Eq.21})), we find that $r_{i0}$ and $r_{0j}$ are the components of local unit Bloch vectors $\hat{n}$ and $\hat{n}^\prime$, respectively (i,j=1,2,3). The matrix elements $r_{ij}$ are the elements of the correlation matrix R which is exactly equal to Eq.(\ref{Eq.25}). Thus, we have incorporated the correlation matrix R inside the Bloch matrix $\textbf{M}_{e\mu}$. Since, we can decompose the Bloch matrix $\textbf{M}_{e\mu}$ in terms of Bloch-vectors components of two sub-systems ($\rho^e_{2\times2}$ and $\rho^\mu_{2\times2}$), therefore, the two-qubit density matrix $\rho^{e\mu}_{4\times4}=\rho^e_{2\times2}\otimes\rho^\mu_{2\times2}$ is a separable state (or product state). In fact, such interpretations of the Bloch matrix $\textbf{M}$ are also valid for the other separable states $\rho^{ee}_{4\times4}=\rho^e_{2\times2}\otimes\rho^e_{2\times2}$, $\rho^{\mu\mu}_{4\times4}=\rho^\mu_{2\times2}\otimes\rho^\mu_{2\times2}$, and $\rho^{\mu e}_{4\times4}=\rho^\mu_{2\times2}\otimes\rho^e_{2\times2}$. 

Furthermore, the concurrence is the measure of entanglement in the two qubit system and it is defined as \cite{Wootters:1997id}
 \begin{eqnarray}
 {C(\rho_{4\times4})= [max(\kappa_1 -\kappa_2 -\kappa_3-\kappa_4 ,0)],}&\nonumber\\
 {\tilde{\rho}_{4\times4}=(\sigma_y\otimes\sigma_y){\rho^*}_{4\times4}(\sigma_y\otimes\sigma_y),}&
 \label{Eq.27}
 \end{eqnarray}
 where $\kappa_1,...,\kappa_4$ are the square roots of the eigenvalues of non-Hermitian matrix $ {\rho}_{4\times4}\tilde{\rho}_{4\times4}$ in decreasing order. The $\tilde{\rho}_{4\times4}$ is the \textquotedblleft {spin-flipped}\textquotedblright density matrix, where the asterisk denotes the complex conjugation in the two qubit standard basis ($\ket{00},\ket{01},\ket{10},\ket{11}$), and $\sigma_x$, $\sigma_y$ are Pauli matrices. We find that for the state $\rho^{e\mu}_{4\times4}=\rho^e_{2\times2}\otimes\rho^\mu_{2\times2}$ (see Eq.(\ref{Eq.22})), all eigenvalues of ${\rho}^{e\mu}_{4\times4}\tilde{\rho}^{e\mu}_{4\times4}$ are zero i.e., $\kappa_1=\kappa_2=\kappa_3=\kappa_4=0$, which mean according to Eq.(\ref{Eq.27}) the concurrence  $C(\rho^{e\mu}_{4\times4})$ is 0. Similarly, for all other possible states: $\rho^{ee}_{4\times4}=\rho^e_{2\times2}\otimes\rho^e_{2\times2}$, $\rho^{\mu\mu}_{4\times4}=\rho^\mu_{2\times2}\otimes\rho^\mu_{2\times2}$, and $\rho^{\mu e}_{4\times4}=\rho^\mu_{2\times2}\otimes\rho^e_{2\times2}$, the concurrence is
 \begin{equation}
 C(\rho^{e\mu}_{4\times4})=C(\rho^{ee}_{4\times4})=C(\rho^{\mu\mu}_{4\times4})=C(\rho^{\mu e}_{4\times4})=0.
 \end{equation}
 We see that concurrence is zero for all the states as expected because they are separable states and we know that there should not be any quantum correlations exist between any two sub systems ($\rho^e_{2\times2}$ and $\rho^\mu_{2\times2}$) of a given separable state. However, if a given state is not separable $\ket{\nu_{e\mu}(\theta,\phi)}\neq\ket{\nu_e(\theta,\phi)}\otimes\ket{\nu_\mu(\theta,\phi)}$ then it is an entangled state. 

 Now, we map the neutrino mass eigenstates $\ket{\nu_1}$ and $\ket{\nu_2}$ directly to the bipartite qubit states as $\ket{\nu_1}=\ket{1}_1\otimes\ket{0}_2$, $\ket{\nu_2}=\ket{0}_1\otimes\ket{1}_2$. In that case, using Eq.(\ref{Eq.8}), the time evolved electron flavour neutrino state $\ket{\nu_e(t)}$ in superposition of two-qubit mass eigenstates, parametrized by $\theta$ and $\phi$, can be written as 
\begin{equation}        
        \ket{\nu_e(\theta,\phi)}={\cos}\theta\ket{10}-{\sin}\theta e^{-i\phi}\ket{01},
        \end{equation}
        and its two-qubit density matrix is
        \begin{equation}
            \rho^{e}_{4\times4}=\ket{\nu_e(\theta,\phi)}\bra{\nu_e(\theta,\phi)}=\begin{pmatrix}
              0& 0 & 0 & 0\\
              0 & {\cos}^2\theta & -{\cos}\theta {\sin}\theta e^{i\phi} & 0\\
              0 & -{\sin}\theta {\cos}\theta e^{-i\phi} & {\sin}^2\theta & 0\\
              0 & 0 & 0 & 0\\
            \end{pmatrix}.
            \label{Eq.30}
        \end{equation}
   Using Eq.(\ref{Eq.30}) in Eq.(\ref{Eq.19}) and in Eq.(\ref{Eq.20}), the Bloch matrix $\textbf{M}$ (see Eq.(\ref{Eq.21})) for $\rho^e_{4\times4}$ is obtained as
     
\begin{equation}
\textbf{M}_e=\left[\begin{array}{c|ccc}

1 & 0 & 0 &  -{\cos}^2\theta+{\sin}^2\theta \\

\hline

0 & -2{\sin}\theta {\cos}\theta {\cos}\phi & -2{\cos}\theta {\sin}\theta {\sin}\phi & 0 \\

0 & 2{\cos}\theta {\sin}\theta {\sin}\phi & -2{\cos}\theta {\sin}\theta {\cos}\phi & 0\\

{\cos}^2\theta-{\sin}^2\theta & 0 & 0 & -{\cos}^2\theta-{\sin}^2\theta\\
\end{array}\right ].
\label{Eq.31}
\end{equation}   
       
         We notice from Eq.(\ref{Eq.31}) that some components of the local Bloch vector for the individual system is zero, and thus we cannot decompose the Bloch matrix $\textbf{M}_{e}$ in terms of Bloch-vectors components of two subsystems. Therefore, the two-qubit density matrix $\rho^{e}_{4\times4}$ of the state $\ket{\nu_e(\theta,\phi)}$ is an entangled state.
         At $\theta=\frac{\pi}{4}$ and $\phi=0$, the Bloch matrix $\textbf{M}_e$ of $\rho^e_{2\times2}$ become 
  \begin{equation}       
\textbf{M}_{\psi^-}=\left[\begin{array}{c|ccc}

1 & 0 & 0 & 0 \\

\hline

0 & -1 & 0 & 0 \\

0 & 0 & -1 & 0\\

0 & 0 & 0 & -1\\
\end{array}\right],
\end{equation}
which is identified as the Bloch matrix of two qubit Bell's state  $\ket{\psi^{-}}=\frac{1}{\sqrt{2}}(\ket{01}-\ket{10})$.
Similarly, for the state $\ket{\nu_\mu(\theta, \phi)}={\sin}\theta\ket{10}+{\cos}\theta e^{-i\phi}\ket{01}$, we have
        \begin{equation}
            \rho^{\mu}_{4\times4}=\ket{\nu_\mu(\theta, \phi)}\bra{\nu_\mu(\theta, \phi)}=\begin{pmatrix}
              0& 0 & 0 & 0\\
              0 & {\sin}^2\theta & {\cos}\theta {\sin}\theta e^{i\phi} & 0\\
              0 & {\sin}\theta {\cos}\theta e^{-i\phi} & {\cos}^2\theta & 0\\
              0 & 0 & 0 & 0\\
            \end{pmatrix},
        \end{equation}
        and the corresponding Bloch Matrix is
        \begin{equation}       
\textbf{M}_{\mu}=\left[\begin{array}{c|ccc}
1& 0 & 0 & -{\cos}^2\theta+{\sin}^2\theta\\

\hline

0 & 2{\sin}\theta {\cos}\theta {\cos}\phi & 2{\cos}\theta {\sin}\theta {\sin}\phi & 0\\
0 & -2{\cos}\theta {\sin}\theta {\sin}\phi & 2{\cos}\theta {\sin}\theta {\cos}\phi & 0\\
{\cos}^2\theta-{\sin}^2\theta & 0 & 0 & -{\cos}^2\theta-{\sin}^2\theta\\
\end{array}\right].
\end{equation}
       At $\theta=\frac{\pi}{4}$ and $\phi=0$, the Bloch-matrix $\textbf{M}_\mu$ become
       
        \begin{equation}
        \textbf{M}_{\psi^+}=\left[\begin{array}{c|ccc}
        1&0&0&0\\
        \hline
        0&1&0&0\\
        0&0&1&0\\
        0&0&0&-1\\
       \end{array}\right],
        \end{equation}
        which is identified as the Bloch matrix of two qubit Bell's state  $\ket{\psi^{+}}=\frac{1}{\sqrt{2}}(\ket{01}+\ket{10})$. Furthermore, concurrence for $\rho^e_{4\times4}$ and $\rho^\mu_{4\times4}$ we get as $C(\rho^e_{4\times4})=C(\rho^\mu_{4\times4})={\sin}2\theta$, which tends to 1 at $\theta=\frac{\pi}{4}$. The nonzero value of concurrence shows that $\rho^e_{4\times4}$ and $\rho^\mu_{4\times4}$ is a bipartite entangled pure state when time evolved neutrino flavour state are linear superposition of mass eigenstates basis. 
        
However, in general, neutrinos change its flavour while traveling in space with time. Therefore, it is useful to quantify concurrence when time evolved neutrino flavour states are linear superposition of flavour basis. In that case, the state of the mass eigenstates can be written in linear superposition of flavour basis as
\begin{equation}
\begin{pmatrix}
\ket{\nu_1}\\
\ket{\nu_2}\\
\end{pmatrix}=U(\theta)\begin{pmatrix}
\ket{\nu_e}\\
\ket{\nu_\mu}\\
\end{pmatrix}.
\label{36}
\end{equation}
Then using Eq.(\ref{36}) in Eq.(\ref{Eq.11}), the time evolved electron and muon flavour neutrino state ($\ket{\nu_e(t)}$ and $\ket{\nu_\mu(t)}$) in linear superposition of flavour basis can be simply written as

 \begin{eqnarray}
                 {\ket{\nu_e (\theta,\phi)}_f=({\cos}^2\theta +{\sin}^2\theta e^{-i\phi})\ket{\nu_e}+{\sin}\theta {\cos}\theta (1-e^{-i\phi })\ket{\nu_\mu},}&\nonumber\\
                 {} \label{37}&\\
                 {\ket{\nu_\mu(\theta,\phi)}_f={\sin}\theta {\cos}\theta(1-e^{-i\phi})\ket{\nu_e}+({\sin}^2\theta+{\cos}^2\theta e^{-i\phi})\ket{\nu_\mu},}&\nonumber
                 \end{eqnarray}

respectively, where suffix f represent that the flavour neutrino state should be written in flavour basis but not in mass basis. $L=ct$ (with $c$ being the speed of light) is the distance traveled by the neutrino particle. $\phi=\frac{\Delta m^2 t}{2E\hbar}$ is typically a function of $L/E$ as neutrino masses are very small so in the ultra-relativistic limit in natural units ($c=1,\hbar=1) $  $L\approx t$. For the state $\ket{\nu_e (\theta,\phi)}_f$, the flavour transition probabilities are 
\begin{eqnarray}
{P_{e\rightarrow e}=|<\nu_e (\theta,\phi)|\nu_e>|^2_f={\cos}^4\theta+{\sin}^4\theta + 2{\sin}^2\theta {\cos}^2\theta {\cos}\phi,\hspace{1.5cm}}&\nonumber\\
{P_{e\rightarrow \mu}=|<\nu_e (\theta,\phi)|\nu_\mu>|^2_f=1-P_{e\rightarrow e}.\hspace{2.3cm}}&
\end{eqnarray}

Now we map flavour states at time t=0 to bipartite state in the two qubit system as $\ket{\nu_e}=\ket{1}_e\otimes\ket{0}_\mu$ and $\ket{\nu_\mu}=\ket{0}_e\otimes\ket{1}_\mu$, therefore Eq.(\ref{37}) become
\begin{eqnarray}
                 {\ket{\nu_e (\theta,\phi)}_f=({\cos}^2\theta +{\sin}^2\theta e^{-i\phi})\ket{10}+{\sin}\theta {\cos}\theta (1-e^{-i\phi })\ket{01},}&\nonumber\\
                 {\ket{\nu_\mu(\theta,\phi)}_f={\sin}\theta {\cos}\theta(1-e^{-i\phi})\ket{10}+({\sin}^2\theta+{\cos}^2\theta e^{-i\phi})\ket{01}.}&\nonumber\\
            \end{eqnarray}
 
The density matrix ${\rho^e_f}_{4\times4}=\ket{\nu_e (\theta,\phi)}_f {}_f\bra{\nu_e (\theta,\phi)}$ of the state $\ket{\nu_e (\theta,\phi)}_f$ in the two-qubit standard basis is

\begin{equation}
{\rho^e_f}_{4\times4}=\begin{pmatrix}
              0& 0 & 0 & 0\\
              0 & |({\cos}^2\theta +{\sin}^2\theta e^{-i\phi})|^2 & ({\cos}^2\theta +{\sin}^2\theta e^{-i\phi}){\sin}\theta {\cos}\theta (1-e^{i\phi}) & 0\\
              0 & ({\cos}^2\theta +{\sin}^2\theta e^{i\phi}){\sin}\theta {\cos}\theta (1-e^{-i\phi}) & |{\sin}\theta {\cos}\theta (1-e^{-i\phi})|^2 & 0\\
              0 & 0 & 0 & 0\\
            \end{pmatrix}.
\end{equation}

We construct the spin-flip density matrix ${\tilde{\rho}^e_{f4\times4}}$ and find that only one eigenvalues of the product state ${\rho^e_f}_{4\times4}{\tilde{\rho}^e_{f4\times4}} $ is nonzero, i.e. $\kappa_1=2\sqrt{P_{e\rightarrow e}P_{e\rightarrow \mu}}$, which means that according to equation Eq.(\ref{Eq.27}), the concurrence for the state ${\rho^e_f}_{4\times4}$ is quantified as  
\begin{equation}
C({\rho^e_f}_{4\times4})=2\sqrt{P_{e\rightarrow e}P_{e\rightarrow \mu}}.
\label{40}
\end{equation}
When $P_{e\rightarrow e}=P_{e\rightarrow \mu}=\frac{1}{2}$, the concurrence $C({\rho^e_f}_{4\times4})$ tends to 1. This results show that time evolved electron flavour neutrino states is a bipartite qubit entangled state. This result is also valid for the state $\ket{\nu_\mu(\theta,\phi)}_f$.  

 \section{SU(3) Poincar\'e sphere for three-flavour neutrinos}
          \label{Sec.4}

 This section uses the Gell-Mann matrices, instead of Pauli's matrices, to map the three flavour neutrino states onto the SU(3) Poincar\'e sphere. In the same way as the mass eigenstates of a two flavour neutrino system were mapped to qubits in the two-dimensional Hilbert space $\mathcal{H}^2$, we now consider the three dimensional Hilbert space $\mathcal{H}^3$. A qutrit is realized by three mutually orthogonal states \cite{Caves:2000}: $\ket{1},\ket{2},\ket{3}$ . 
 
 A quantum state in the Hilbert space $\mathcal{H}^3$ spanned by the three orthogonal qutrit states $\ket{1}=\begin{pmatrix}
            1\\
            0\\
            0\\
            \end{pmatrix};$ $\ket{2}=\begin{pmatrix}
            0\\
            1\\
            0\\
            \end{pmatrix};$ and $\ket{3}=\begin{pmatrix}
            0\\
            0\\
            1\\
            \end{pmatrix}$ is $\ket{\psi}=\alpha\ket{1}+\beta\ket{2}+\gamma\ket{3}$ where $ |\alpha|^2+|\beta|^2+|\gamma|^2=1$. Using the polar representation a quantum state $\ket{\psi}$ does not change if multiplied by an overall phase, the equivalent quantum state is ($0\leq\theta,\phi\leq \frac{\pi}{2}; 0\leq\xi_1,\xi_2<2\pi$) \cite{Caves:2000} 
            \begin{equation}
            \ket{\psi}=e^{i\xi_1}{\sin}(\theta){\cos}(\phi)\ket{1}+e^{i\xi_2}{\sin}(\theta){\sin}(\phi)\ket{2}+{\cos}(\theta)\ket{3}.
            \label{Eq.36}
            \end{equation}
           The corresponding density matrix for the state $\ket{\psi}$ is
\begin{equation}
\rho_{3\times3}(\psi)=\ket{\psi}\bra{\psi}= \begin{pmatrix}
{\sin}^2\theta {\cos}^2\phi & \frac{e^{i(\xi_1-\xi_2)}}{2} {\sin}^2\theta {\sin}(2\phi) & \frac{e^{i\xi_1}}{2} {\sin}(2\theta){\cos}\phi\\
\frac{e^{i(\xi_2-\xi_1)}}{2} {\sin}^2\theta {\sin}(2\phi) & {\sin}^2\theta {\sin}^2\phi & \frac{e^{i\xi_2}}{2} {\sin}(2\theta){\sin}\phi\\
\frac{e^{-i\xi_1}}{2} {\sin}(2\theta){\cos}\phi & \frac{e^{-i\xi_2}}{2} {\sin}(2\theta){\sin}\phi & {\cos}^2{\theta} \\
\end{pmatrix}.
\end{equation}            
            
  The pure state $\ket{\psi}$ in Eq.(\ref{Eq.36}) is dependent on 4 parameters $\theta,\phi,\xi_1,\xi_2$. However, the five parameters are needed to characterize the neutrino state, and the sum of the squares of the state's coefficients should be one.

            Now we will define the density operator of a qutrit system using SU(3) in general and then map it to the neutrino system. The density matrix $\rho_{3\times3}=\ket{\psi}\bra{\psi}$ is a $3\times3$ unitary matrix such that ${(\rho_{3\times3})}^\dagger=\rho_{3\times3} $; and $Tr(\rho_{3\times3})=1$. The qutrit representation of the density matrix uses the eight (Hermitian, traceless) generators of SU(3) as an operator basis called the Gell-Mann matrices \cite{Caves:2000}. By supplementing the eight Gell-Mann matrices $\lambda_i, i=1,...,8$  with the unit operator $\lambda_0\equiv\sqrt{\frac{2}{3}}1$, the qutrit density matrix operator is a vector in the space spanned by $\lambda_\alpha, \alpha=0,...,8$ and therefore can be written as
\begin{eqnarray}
{\rho_{3\times3}=\ket{\psi}\bra{\psi}=\frac{1}{3}c_{\alpha}\lambda_{\alpha}\hspace{2cm}}&\nonumber\\
{=(\alpha\ket{1}+\beta\ket{2}+\gamma\ket{3})(\alpha^*\bra{1}+\beta^*\bra{2}+\gamma^*\bra{3})}&,
\end{eqnarray}
\begin{eqnarray}
{= \sqrt{\frac{3}{2}}\lambda_0+ \frac{1}{2}(\alpha\beta^*+\beta\alpha^*)\lambda_1 + \frac{i}{2}(\alpha\beta^*-\beta\alpha^*)\lambda_2+\frac{1}{2}(|\alpha|^2 - |\beta|^2)\lambda_3 + \frac{1}{2}(\alpha\gamma^*+\gamma\alpha^*)\lambda_4}&\nonumber\\
    {+\frac{i}{2}(\alpha\gamma^*-\gamma\alpha^*)\lambda_5+\frac{1}{2}(\beta\gamma^*+\gamma\beta^*)\lambda_6+\frac{i}{2}(\beta\gamma^*-\gamma\beta^*)\lambda_7+\frac{1}{2\sqrt{3}}(|\beta|^2-2|\gamma|^2)\lambda_8,}&
\end{eqnarray}
 the (real) expansion coefficients are 
\begin{equation}
c_{\alpha}=\frac{3}{2}tr(\rho_{3\times3} \lambda_{\alpha}).
\end{equation}
Normalization implies that $c_{0}=\sqrt{\frac{3}{2}}$, so the density operator can be simplified to the $SU(3)$ equivalent of Eq.(\ref{Eq.5}), which we shall show is the Poincar\'e sphere representation of the qutrit states
\begin{equation}
\rho_{3\times3}=\frac{1}{3}
(1+c_j\lambda_j)=\frac{1}{3}(1+\Vec{c}.\Vec{\lambda}),
\end{equation}
 where $\Vec{c}=c_j\hat{e_j}$ and $\Vec{\lambda}=\lambda_j\hat{e_j}$.
 
To find the coefficients $c_i$ we note that 
\begin{equation}
\rho_{3\times3}^2=\frac{1}{9}(1+\frac{2}{3} \Vec{c}.\Vec{c})1+\frac{1}{3}\Vec{\lambda}.(\frac{2}{3}\Vec{c}+\frac{1}{3\sqrt{3}}\Vec{c}\star\Vec{c}),
\end{equation}
where the ``star'' product is defined as
\begin{equation}
\Vec{c}\star\Vec{d}\equiv \hat{e_j}d_{jkl}c_kd_l.
\end{equation}

The characterization of the Gell-Mann matrices and $d_{jkl}$ can be found in \cite{Lichtenberg} or any group theory text book. The star product condition is well explained in ref.\cite{Caves:2000}. For a pure state $\ket{\psi}$, $\rho_{3\times3}^2=\rho_{3\times3}$, so we must have $\Vec{c}.\Vec{c}=3$ and $\Vec{c}\star\Vec{c}=\sqrt{3}\Vec{c}$. Defining the eight dimensional unit vector $\hat{n}=\Vec{c}/\sqrt{3}$, we find any qutrit pure state density matrix can be written as
\begin{equation}
\rho_{3\times3}=\ket{\psi}\bra{\psi}=\frac{1}{3}(I+\sqrt{3}\hat{n}.\Vec{\lambda}),
\label{Eq.47}
\end{equation}
 where $\hat{n}$ satisfies
\begin{equation}
\hat{n}.\hat{n}=1\; and \; \hat{n}\star\hat{n}=\hat{n}.
\label{Eq.48}
\end{equation} 
Eq.(\ref{Eq.47}) is the equation for the 7-dimensional unit sphere  $\hat{n}\in S^7$ embedded in Euclidean eight dimensional space $\mathcal{R}^{8}$ spanned by the Gell-Mann matrices. It is in fact a representation of the coset space SU(3)/U(2) \cite{Arvind:1996rj}, with components of unit vector $\hat{n}$ given by
\begin{equation}
n_j=\frac{\sqrt{3}}{2}tr(\rho_{3\times3}\lambda_j)=\frac{\sqrt{3}}{2}\bra{\psi}\lambda_j\ket{\psi}.
\label{Eq.49}
\end{equation}
 Thus, we have outlined the Poincar\'e sphere representation of the density matrix in $\mathcal{H}^3$.
 
 Three-flavour neutrino oscillations involve a Hilbert space $\mathcal{H}^3$ and the mixing matrix is given by the SU(3) matrix \cite{Carlo:2010}. Let the mass eigenstates of the three-flavour neutrino system be $\ket{\nu_1}$, $\ket{\nu_2}$ and $\ket{\nu_3}$ then the relationship between the mass eigenstates and the flavour states is
\begin{equation}
        \begin{pmatrix}
        \ket{\nu_e}\\
        \ket{\nu_\mu}\\
        \ket{\nu_\tau}\\
        \end{pmatrix}=U^*(\theta,\phi,\eta,\delta_{CP})\begin{pmatrix}
        \ket{\nu_1}\\
        \ket{\nu_2}\\
        \ket{\nu_3}\\
        \end{pmatrix},
        \label{Eq.50}
    \end{equation}
where $U(\theta,\phi,\eta,\delta_{CP})$ is the Unitary PMNS (Pontecorvo-Maki-Nakagawa-Sakata) neutrino mixing matrix 
\begin{equation} 
U(\theta,\phi,\eta,\delta_{CP})=
 \begin{pmatrix} 
 C_{\theta} C_{\phi} & S_{\theta} C_{\phi} & S_{\phi}e^{-i\delta_{CP}}\\
 -S_{\theta} C_{\eta}-C_{\theta} S_{\phi} S_{\eta}e^{i\delta_{CP}} & C_{\theta} C_{\eta} -S_{\theta} S_{\phi} S_{\eta}e^{i\delta_{CP}}& C_{\phi} S_{\eta}\\
 S_{\theta} S_{\eta} -C_{\theta} S_{\phi} C_{\eta}e^{i\delta_{CP}}   & -C_{\theta} S_{\eta} -S_{\theta} S_{\phi}C_{\eta}e^{i\delta_{CP}}  & C_{\phi} C_{\eta} \\ 
 \end{pmatrix}\in SU(3),
 \label{Eq.51}
 \end{equation}
 where $(S_\theta,S_\phi,S_\eta)\equiv ({\sin}\theta_{12},{\sin}\theta_{13},{\sin}\theta_{23})$; $(C_\theta,C_\phi,C_\eta)\equiv ({\cos}\theta_{12},{\cos}\theta_{13},{\cos}\theta_{23})$, the $\theta_{ij}$'s are the neutrino mixing angles between the states $i$ and $j$ $(i,j=1,2,3)$ \cite{Giganti:2017fhf}. 
            
The three-flavour states of a neutrino system can be written in the qutrit basis by identifying the mass eigenstates with the qutrit basis states of the three-dimensional Hilbert space $\mathcal{H}^3$ as
\begin{equation}
            \ket{1}=\ket{\nu_1}\;\; ;
             \ket{2}=\ket{\nu_2}\;\; ;\ket{3}=\ket{\nu_3}.
             \label{Eq.52}
            \end{equation}
Without loss of generality, we take $\delta_{CP}=0$ and write the time evolved electron flavor neutrino state as
        \begin{equation}
    \ket{\nu_e (t)}=e^{-iE_1t/\hbar} C_{\theta} C_{\phi}\ket{1}+e^{-iE_2t/\hbar} (-S_{\theta} C_{\eta}-C_{\theta} S_{\phi} S_{\eta}) \ket{2}+ e^{-iE_3t/\hbar} (S_{\theta} S_{\eta} -C_{\theta} S_{\phi} C_{\eta} ) \ket{3}
    \end{equation}
Similarly, the time evolved  $\ket{\nu_{\mu}(t)}$ and $\ket{\nu_{\tau}(t)}$ neutrino flavor states can be written as
\begin{eqnarray}
    {\ket{\nu_\mu (t)}=e^{-iE_1t/\hbar} S_{\theta} C_{\phi}\ket{1}+e^{-iE_2t/\hbar} (C_{\theta} C_{\eta} -S_{\theta} S_{\phi} S_{\eta}) \ket{2}+ e^{-iE_3t/\hbar}( -C_{\theta} S_{\eta} -S_{\theta} S_{\phi}C_{\eta} ) \ket{3}}\label{Eq.54}&\\
    {\textit{and}\hspace{0.2cm}\ket{\nu_\tau (t)}=e^{-iE_1t/\hbar} S_{\phi} \ket{1}+e^{-iE_2t/\hbar} C_{\phi} S_{\eta} \ket{2}+ e^{-iE_3t/\hbar} C_{\phi} C_{\eta} \ket{3}}&
    \end{eqnarray}
respectively. Taking the ultra-relativistic limit $L\approx t$ ($c=1$,$\hbar=1$) and defining $\xi_1=(E_3-E_1)t/\hbar\approx\Delta m^2_{31}L/2E$, and $\xi_2=(E_3-E_2)t/\hbar\approx\Delta m^2_{32}L/2E$, the normalized time evolved electron neutrino flavour state $\ket{\nu_e(t)}$ in qutrit basis, parametrized by three different mixing angle $\theta$, $\phi$, $\eta$ and with two arbitrary phases $\xi_1$ and $\xi_2$ ($0\leq\theta,\phi,\eta\leq \frac{\pi}{2}; 0\leq\xi_1,\xi_2<2\pi$) can be written as
            \begin{equation}
                \ket{\nu_e(\theta,\phi,\eta,\xi_1,\xi_2)}=e^{i\xi_1}C_\theta C_\phi \ket{1}+ e^{i\xi_2} (-S_\theta C_\eta - C_\theta S_\phi S_\eta) \ket{2}+( S_\theta S_\eta- C_\theta S_\phi C_\eta)\ket{3}.
                \label{Eq.56}
            \end{equation}
  
 The $3\times3$ density matrix of the state $\ket{\nu_e(\theta,\phi,\eta,\xi_1,\xi_2)}$ is
\begin{equation}
       \rho^e_{3\times3} = \ket{\nu_e(\theta,\phi,\eta,\xi_1,\xi_2)}\bra{\nu_e(\theta,\phi,\eta,\xi_1,\xi_2)},
       \label{Eq.57} 
       \end{equation}  
       which in matrix form is

       \begin{equation}
        \rho^e_{3\times3} =\begin{tiny}\begin{pmatrix}
       C^2_\theta C^2_\phi & -e^{-i(\xi_2-\xi_1)} C_\theta C_\phi (S_\theta C_\eta+C_\theta S_\phi S_\eta) &  e^{i\xi_1} C_\theta C_\phi  (S_\theta S_\eta-C_\theta C_\phi C_\eta)\\   
                
       -e^{i(\xi_2-\xi_1)}  C_\theta C_\phi(S_\theta C_\eta+C_\theta S_\phi S_\eta) & {(S_\theta C_\eta+C_\theta S_\phi S_\eta})^2 & -e^{i\xi_2}(S_\theta C_\eta+C_\theta S_\phi S_\eta)(S_\theta S_\eta-C_\theta C_\phi C_\eta) \\       
         
   e^{-i\xi_1}  C_\theta C_\phi  (S_\theta S_\eta - C_\theta C_\phi C_\eta) &-e^{-i\xi_2}(S_\theta C_\eta+C_\theta S_\phi S_\eta)(S_\theta S_\eta-C_\theta C_\phi C_\eta)& (S_\theta S_\eta -C_\theta C_\phi C_\eta)^2 \\      
       \end{pmatrix}.\nonumber   
       \end{tiny}      
            \end{equation}
         
The density matrix $\rho_{3\times3}^{e}$ satisfies the relation ${(\rho^{e}_{3\times3})}^\dagger=(\rho^e_{3\times3})^2=\rho^e_{3\times3}$; and $Tr(\rho^e_{3\times3})=1$. The density matrix for the time evolved electron flavour neutrino state $\ket{\nu_{e}(\theta,\phi,\eta,\xi_1,\xi_2)}$ can now be cast into the form
\begin{equation}
\rho^e_{3\times3} = \ket{\nu_e(\theta,\phi,\eta,\xi_1,\xi_2)}\bra{\nu_e(\theta,\phi,\eta,\xi_1,\xi_2)}=\frac{1}{3}(I+\sqrt{3}\hat{n}.\Vec{\lambda}).
\label{Eq.58}
\end{equation}
 The unit vector  ($\hat{n}.\hat{n}=1$), in the Euclidean eight dimensional space $\mathcal{R}^{8}$ is 
\begin{equation}
\hat{n}(\theta,\phi,\eta,\xi_1,\xi_2)=n_1\hat{e_1}+n_2\hat{e_2}+n_3\hat{e_3}+n_4\hat{e_4}+n_5\hat{e_5}+n_6\hat{e_6}+n_7\hat{e_7}+n_8\hat{e_8}.
\label{59}
\end{equation}
 Using the density matrix form ($\rho^e_{3\times3}$) of Eq.(\ref{Eq.57}) in Eq.(\ref{Eq.49}), the components of the unit vector $\hat{n}(\theta,\phi,\eta,\xi_1,\xi_2)$ can be obtained as:
 \begin{eqnarray}
          {n_1=-\sqrt{3}C_\theta C_\phi (S_\theta C_\eta+C_\theta S_\phi S_\eta) {\cos}(\xi_2-\xi_1);}&\nonumber\\
          {n_2=-\sqrt{3} C_\theta C_\phi(S_\theta C_\eta+C_\theta S_\phi S_\eta) {\sin}(\xi_2-\xi_1);}&\nonumber\\
           {n_3=\frac{\sqrt{3}}{2}[C^2_\theta C^2_\phi-(S_\theta C_\eta+C_\theta S_\phi S_\eta)^2];}&\nonumber\\
           {n_4=\sqrt{3}C_\theta C_\phi (S_\theta S_\eta-C_\theta C_\phi C_\eta) {\cos}\xi_1 ;}&\nonumber\\
          {n_5=-\sqrt{3}C_\theta C_\phi (S_\theta S_\eta-C_\theta C_\phi C_\eta) {\sin}\xi_1 ;}&\nonumber\\
           {n_6=-\sqrt{3}(S_\theta C_\eta +C_\theta S_\phi S_\eta)(S_\theta S_\eta - C_\theta C_\phi C_\eta) {\cos}\xi_2 ;}&\nonumber\\
            {n_7=\sqrt{3}(S_\theta C_\eta +C_\theta S_\phi S_\eta)(S_\theta S_\eta -C_\theta C_\phi C_\eta) {\sin}\xi_2 ;}&\nonumber\\
            {n_8=\frac{1}{2}[C^2_\theta C^2_\phi +(S_\theta C_\eta +C_\theta S_\phi S_\eta)^2}-2(S_\theta S_\eta -C_\theta C_\phi C_\eta)^2].&\nonumber\\
            \label{Eq.61}
        \end{eqnarray}

The result shows that the time evolved electron flavour neutrino state lies on the $S^7$ sphere in the eight dimensional real vector spaces. Not all the operators on the unit-sphere are pure state, so the star product condition $\hat{n}\star\hat{n}=\hat{n}$ (see Eq.(\ref{Eq.48})) imposes three constraints on the unit vector $\hat{n}(\theta,\phi,\eta,\xi_1,\xi_2)$ (see Eq.(\ref{59})) and therefore reduces the number of arbitrary parameters for the neutrino states. The three constraints give us three orthonormal components of $\hat{n}(\theta,\phi,\eta,\xi_1,\xi_2)$. In the following table, we list the three constraints
and their corresponding orthonormal unit vectors.

\begin{table}[H]
\centering
\begin{tabular}{ |p{0.1\linewidth}|p{0.3\linewidth}|p{0.3\linewidth}|}
 \hline
  S.No. &  Constraints & Corresponding $\hat{n}$   \\
\hline
1. &$\theta=\phi=0$, $\eta=\frac{\pi}{2}$, $\xi_1$ and $\xi_2$ are arbitrary  & $\hat{n}_1=\frac{\sqrt{3}}{2}\hat{e}_3+\frac{1}{2}\hat{e}_8$ \\
 \hline
2. & $\theta=\pi/2,\phi=\eta=0$, $\xi_1$ and $\xi_2$ are arbitrary &$\hat{n}_2=-\frac{\sqrt{3}}{2}\hat{e}_3+\frac{1}{2}\hat{e}_8$   \\
 \hline
3. & $\theta=\phi=\eta=\pi/2$, $\xi_1$ and $\xi_2$ are arbitrary & $\hat{n}_3=-\hat{e}_8$ \\
 \hline
\end{tabular}
\caption{ The three constraints coming from star product condition $\hat{n}\star\hat{n}=\hat{n}$ (see Eq.(\ref{Eq.48})) and their corresponding orthonormal unit vectors.}
\label{table-1}
\end{table}

     These orthonormal states also satisfies the condition \cite{Arvind:1996rj}
\begin{eqnarray}
{|<\psi|\psi^{\prime}>|^2=tr(\rho\rho^{\prime})=\frac{1}{3}(1+2\hat{n}.\hat{n}^{\prime}),}&\nonumber\\
{0\leq tr(\rho\rho^{\prime})\leq 1 \Longleftrightarrow 0\leq {\cos}^{-1}(\hat{n}.\hat{n}^{\prime})\leq\frac{2\pi}{3}.}&
\label{Eq.62}
\end{eqnarray}   

  We find that the angle formed between any two unit vectors  ($\hat{n_1}$, $\hat{n_2}$, $\hat{n_3}$) is $\frac{2\pi}{3}$, since ${\cos}^{-1}(\hat{n}_1.\hat{n}_2)={\cos}^{-1}(\hat{n}_1.\hat{n}_3)={\cos}^{-1}(\hat{n}_2.\hat{n}_3)={\cos}^{-1}(\frac{-1}{2})=\frac{2\pi}{3}$. Identifying the three orthonormal basis of qutrit as the mass eigenstates of neutrinos (see Eq.(\ref{Eq.52})), the Eq.(\ref{Eq.62}) shows that the pure state $\ket{\nu_e(\theta,\phi,\eta,\xi_1,\xi_2}$ (see Eq.(\ref{Eq.56})) in an orthonormal basis ($\ket{\nu_1}$, $\ket{\nu_2}$, $\ket{\nu_3}$) has unit vectors ($\hat{n_1}$, $\hat{n_2}$, $\hat{n_3}$) that lie in a plane at the vertices of an equilateral triangle which we term as a ``qutrit triangle''. If one takes the three canonical basis vectors of $\mathcal{H}^3$ as usual, the three vertices of an equilateral triangle are 
                \begin{eqnarray}
                    {(n_3,n_8)_A=(\frac{\sqrt{3}}{2},\frac{1}{2})\longrightarrow (1,0,0)^T=\ket{1}=\ket{\nu_1};\hspace{1cm}}&\\
                    {(n_3,n_8)_B=(-\frac{\sqrt{3}}{2},\frac{1}{2})\longrightarrow (0,1,0)^T=\ket{2}=\ket{\nu_2};\hspace{1cm}}&\\
                    {(n_3,n_8)_C=(0,-1)\longrightarrow (0,0,1)^T=\ket{3}=\ket{\nu_3};\hspace{1cm}}&
                \end{eqnarray} 
                which are identified with the generalized W-states of neutrinos \cite{ Jha:2020hyh}. Thus, we generalize the concept of tripartite mode entanglement by considering neutrinos as qutrits.         
            \begin{figure}[H] 
        \centering
         \includegraphics[width=0.8\columnwidth]{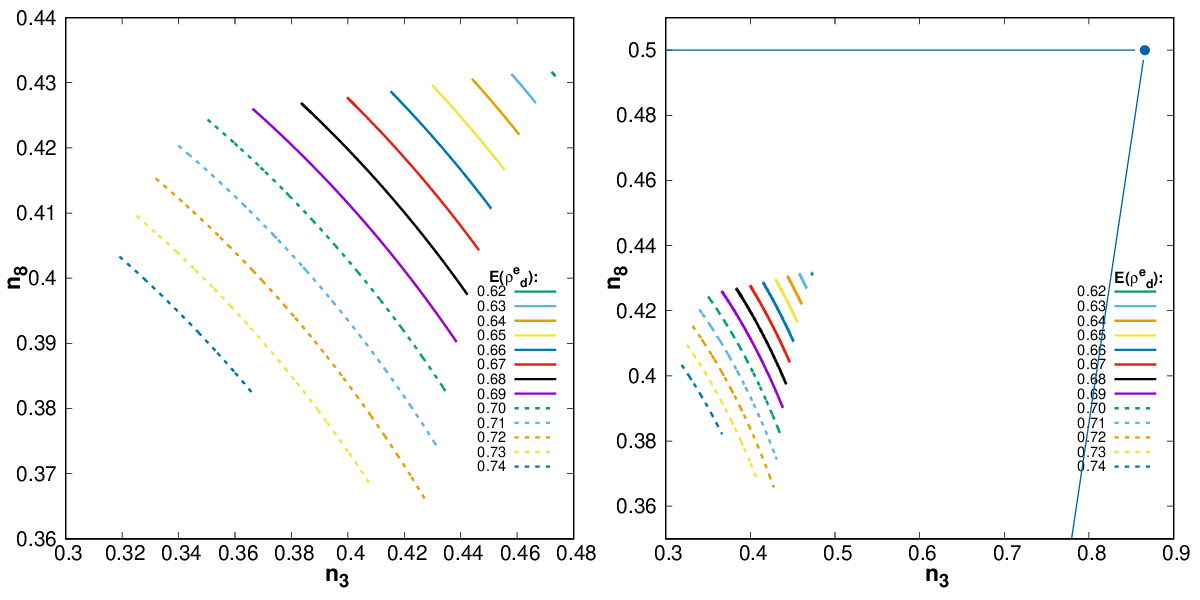}    
       \caption{In Fig.(a) the equi-mixing curves of $E(\rho^e_d)$ in the $n_3$ and $n_8$ plane is shown using the current experimental bounds of the $3\sigma$ range of neutrino parameters \cite{Esteban:2020cvm}. Fig.(b) shows the equi-mixing curves of $E(\rho^e_d)$ in the $n_3$ and $n_8$ plane inside the qutrit triangle.}
\label{fig:1}
\end{figure}    
      
\begin{figure}[H] 
        \centering \includegraphics[width=0.6\columnwidth]{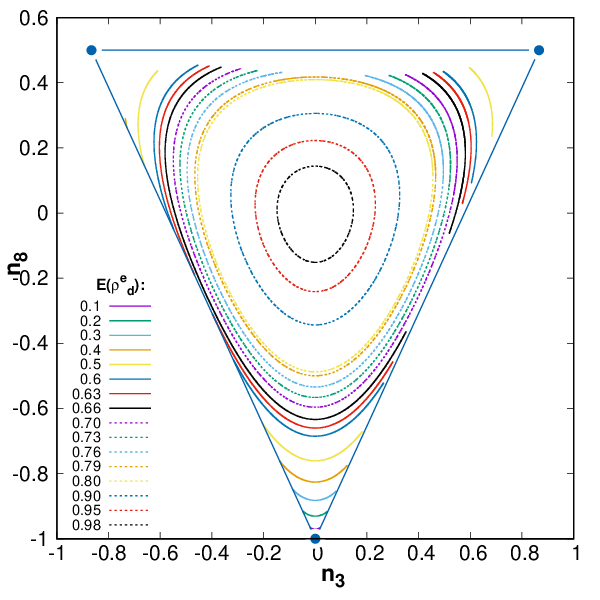}    
        \caption{The equi-mixing curves of $E(\rho^e_d)$ is shown in the $n_3$ and $n_8$ plane when $\theta$ and $\eta$ are vary from 0 to $\pi/2$.}
\label{fig:2}
\end{figure}                         
The diagonal density matrix in the orthonormal basis is the triangle operator, or interior \cite{Caves:2000,Bolukbasi:2006}. We map the neutrino state density matrix $\rho^e_{3\times3}$ of SU(3) space directly to the $\lambda_3$ and $\lambda_8$ basis (two diagonal Gell-Mann matrices) to construct a mixed state density matrix $\rho^e_d{_{(3\times3)}}$. Thus, the density matrix $\rho^e_{3\times3}$ of Eq.(\ref{Eq.58}) is now reduced to a mixed state as
\begin{eqnarray}
                {\rho^e_d{_{(3\times3)}}= \frac{1}{3}(I+\sqrt{3}(n_3\lambda_3+n_8\lambda_8))\hspace{2.5cm}}&\nonumber\\
                   { =\frac{1}{3}\begin{pmatrix}
                    1+\sqrt{3}n_3+n_8 & 0 &0\\
                    0&1-\sqrt{3}n_3+n_8&0\\
                    0&0&1-2n_8&\\
                    \end{pmatrix},\hspace{0.5cm}}&
                \end{eqnarray}  
where $x_1=\frac{1}{3}(1+\sqrt{3}n_3+n_8),x_2=\frac{1}{3}(1-\sqrt{3}n_3+n_8)$, and $x_3=\frac{1}{3}(1-2n_8)$ are three eigenvalues of $\rho^e_d{_{(3\times3)}}$ in terms of $n_3$ and $n_8$. The value of $n_3$ and $n_8$ are given in Eq.(\ref{Eq.61}). We calculate the entropy of mixing of the mixed state $\rho^e_d{_{(3\times3)}}$ by using the formula \cite{Bolukbasi:2006}: 
             \begin{eqnarray}
                 E(\rho^e_d{_{(3\times3)}})=-x_1log_3(x_1)-x_2log_3(x_2)-x_3log_3(x_3).\hspace{0.8cm}
             \end{eqnarray}

 In Fig.(\ref{fig:1}) (Fig.(a) and Fig.(b)), we plot the equi-mixing curves of $E(\rho^e_d)$ in the $n_3$ and $n_8$ plane. We vary $\theta\equiv\theta_{12}$ and $\eta\equiv\theta_{23}$ over $3\sigma$ range of current experimental bounds \cite{Esteban:2020cvm}, we find that neutrinos are inside qutrit triangle for the range of entropy of mixing $E(\rho^e_d)$ approximately between 0.62 to 0.74. In Fig.(\ref{fig:2}), we vary $\theta$ and $\eta$ from 0 to $\pi/2$, and we see that this put the constraints on $\theta_{12}$ and $\theta_{23}$ to be greater than 23 degrees for physical result. 

Furthermore, the SU(2) Poincar\'e sphere representation for two-flavour neutrino oscillations can be deduced from the three-flavour Poincar\'e sphere in SU(3) by imposing the conditions 
\begin{equation}
{{\tan}\theta_{23}}\hspace{0.15cm} {{\tan}\theta_{12}}={\sin}\theta_{13}.
\end{equation}   

In the limit when the mixing between 2 and 3 (49 degrees), and 1 and 2 (33.44 degrees) is greater than the mixing between 1 and 3, $\theta_{13}\approx 0$ (8.57 degrees, ${\sin}\theta_{13}\approx0.15)$. We set $\theta_{13}\approx0$, so that Eq.(\ref{Eq.56}) $\ket{\nu_e(\theta,\phi,\eta,\xi_1,\xi_2)}$ can be reduced to
\begin{equation}
\ket{\nu_e(\theta,\xi_1,\xi_2)}=\begin{pmatrix}
e^{i\xi_1}{\cos}\theta\\
-e^{i\xi_2}{\sin}\theta\\
0\\
\end{pmatrix}.
\end{equation}
where, $0\leq\theta\leq\frac{\pi}{2},\hspace{0.15cm} 0\leq\xi_1,\xi_2<2\pi$. We calculate the density matrix $\rho^e_{3\times 3}=\ket{\nu_e(\theta,\xi_1,\xi_2)}$ $\bra{\nu_e(\theta,\xi_1,\xi_2)}$ of the above reduced state $\ket{\nu_e(\theta,\xi_1,\xi_2)}$ and use it in Eq.(\ref{Eq.49}). We find that the unit vector $\hat{n}$ in eight-dimensional real vector space now reduces to only four non-vanishing components
\begin{equation}
n_1=-\frac{\sqrt{3}}{2} {\sin}2\theta {\cos}(\xi_2-\xi_1);\hspace{0.3cm} n_2=\frac{\sqrt{3}}{2} {\sin}2\theta {\sin}(\xi_2-\xi_1);\hspace{0.3cm}n_3= \frac{\sqrt{3}}{2}{\cos}2\theta;\hspace{0.3cm} n_8=\frac{1}{2},
\end{equation}
 else all are zero, i.e. $n_4=n_5=n_6=n_7=0$. Hence, when there is a hierarchy of mixing between the three states 1, 2 and 3 with the third state almost decoupled (small mixing angle), we retrieve the SU(2) Poincar\'e sphere from the SU(3) Poincar\'e sphere.

    So far, we have considered the Poincar$e^\prime$ sphere representation of a time evolved electron-neutrino flavour state. For completeness, we give the Poincar$e^\prime$ sphere representation of a time evolved muon-neutrino flavour state. The state $\ket{\nu_\mu (t)}$ (see Eq.(\ref{Eq.54})) parametrized by $\theta,\phi,\eta,\xi_1,\xi_2$ in the qutrit basis can be written as
\begin{equation}
    \ket{\nu_{\mu}(\theta,\phi,\eta,\xi_1,\xi_2)}=e^{i\xi_1} S_{\theta} C_{\phi}\ket{1}+e^{i\xi_2} (C_{\theta} C_{\eta} -S_{\theta} S_{\phi} S_{\eta}) \ket{2}+ ( -C_{\theta} S_{\eta} -S_{\theta} S_{\phi}C_{\eta} ) \ket{3},
    \end{equation}    
   and its density matrix $\rho^\mu_{3\times3}$ $= \ket{\nu_\mu(\theta,\phi,\eta,\xi_1,\xi_2)}\bra{\nu_\mu(\theta,\phi,\eta,\xi_1,\xi_2)}$ is
      \begin{equation}
      \rho^\mu_{3\times3} =\begin{tiny}
      \begin{pmatrix}
      S^2_\theta C^2_\phi & e^{-i(\xi_2-\xi_1)} S_\theta C_\phi (C_\theta S_\eta-S_\theta S_\phi S_\eta) &  e^{i\xi_1} S_\theta C_\phi  (-C_\theta S_\eta-S_\theta S_\phi S_\eta)\\   
                
      e^{i(\xi_2-\xi_1)}  S_\theta C_\phi(C_\theta S_\eta - S_\theta S_\phi S_\eta) & {(C_\theta S_\eta-S_\theta S_\phi S_\eta})^2 &  e^{i\xi_2}(C_\theta S_\eta-S_\theta S_\phi S_\eta)(-C_\theta S_\eta-S_\theta S_\phi S_\eta) \\       
         
   e^{-i\xi_1}  S_\theta C_\phi  (-C_\theta S_\eta - S_\theta S_\phi S_\eta) & e^{-i\xi_2}(-C_\theta S_\eta-S_\theta S_\phi S_\eta)(C_\theta S_\eta-S_\theta S_\phi S_\eta)& (-C_\theta S_\eta -S_\theta S_\phi S_\eta)^2 \\      
      \end{pmatrix}.
      \end{tiny}
       \label{Eq.72}
            \end{equation}
    The density matrix $\rho^\mu_{3\times3}$ can be expanded in the Gell-Mann basis as    
     \begin{equation}
     \rho^\mu_{3\times3}=\frac{1}{3}(I+\sqrt{3}\hat{n^\prime}.\Vec{\lambda}),
\end{equation}                  
  
 and by using Eq.(\ref{Eq.72}) in Eq.(\ref{Eq.49}), we get the components of the unit vector $\hat{n^\prime}(\theta,\phi,\eta,\xi_1,\xi_2)$ as
 \begin{eqnarray}
          {n^\prime_1=\sqrt{3}S_\theta C_\phi (C_\theta S_\eta-S_\theta S_\phi S_\eta) {\cos}(\xi_2-\xi_1);}&\nonumber\\
          {n^\prime_2=\sqrt{3} S_\theta C_\phi(C_\theta S_\eta-S_\theta S_\phi S_\eta) {\sin}(\xi_2-\xi_1);}&\nonumber\\
           {n^\prime_3=\frac{\sqrt{3}}{2}[S^2_\theta C^2_\phi-(C_\theta S_\eta-S_\theta S_\phi S_\eta)^2];}&\nonumber\\
           {n^\prime_4=\sqrt{3}S_\theta C_\phi (-C_\theta S_\eta-S_\theta S_\phi S_\eta) {\cos}\xi_1 ;}&\nonumber\\
          {n^\prime_5=-\sqrt{3}S_\theta C_\phi (-C_\theta S_\eta-S_\theta S_\phi S_\eta) {\sin}\xi_1 ;}&\nonumber\\
           {n^\prime_6=\sqrt{3}(C_\theta S_\eta -S_\theta S_\phi S_\eta)(-C_\theta S_\eta - S_\theta S_\phi S_\eta) {\cos}\xi_2 ;}&\nonumber\\
            {n^\prime_7=-\sqrt{3}(C_\theta S_\eta -S_\theta S_\phi S_\eta)(-C_\theta S_\eta -S_\theta S_\phi S_\eta) {\sin}\xi_2 ;}&\nonumber\\
           { n^\prime_8=\frac{1}{2}[S^2_\theta C^2_\phi +(C_\theta S_\eta -S_\theta S_\phi S_\eta)^2}-2(C_\theta S_\eta -S_\theta S_\phi S_\eta)^2].&\nonumber\\
        \end{eqnarray}

\section{Two qutrits flavour neutrino states and generalized concurrence}
\label{Sec.5}
                                                                                                                                                                                                                                                                                                                                                                                                                                                                                                                                                                                                                                                                                                                                                                                                                                                                                                                                                                                                                                                                                                                                                                                                                                                                                                                                                                                                                                                                                                                                                                                                                                                                                                                                                                                                                                                                                                                                                                                                                                                                                                                                                                                                                                                                                                                                                                                    
In general, any two qutrits state is defined as the tensor product of two three dimensional Hilbert spaces, i.e. $\mathcal{H}^3\otimes \mathcal{H}^3$. This section represents a two-qutrit density matrix of the neutrino system based on Gell-Mann matrix tensor products, with the coefficients constituting a generalized matrix analogous to a two-qubit Bloch vector of neutrinos. 

According to Eqs.(\ref{Eq.50}, \ref{Eq.51} and \ref{Eq.52}), in the three neutrino system, in general the time evolved neutrino flavour states in qutrit basis ($\ket{1},\ket{2},\ket{3}$) for the two different sub-system A and B can be represented as ($A,B=e,\mu,\tau$) :
   
    \begin{eqnarray}
      {\ket{\nu_A(\theta,\phi,\eta,\xi_1,\xi_2)}=\alpha_1\ket{1}+\alpha_2\ket{2}+\alpha_3\ket{3}}&\\
      { \ket{\nu_{B}(\theta,\phi,\eta,\xi_1,\xi_2))}=\alpha^\prime_1\ket{1}+\alpha^\prime_2\ket{2}+\alpha^\prime_3\ket{3}}&\nonumber
      \end{eqnarray} 
     where $\ket{\nu_A(\theta,\phi,\eta,\xi_1,\xi_2)}\in\mathcal{H}^3_A$  and $\ket{\nu_B(\theta,\phi,\eta,\xi_1,\xi_2)}\in\mathcal{H}^3_B$, and $|\alpha_1|^2+|\alpha_2|^2+|\alpha_3|^2=1$ and $|\alpha^\prime_1|^2+|\alpha^\prime_2|^2 +|\alpha^\prime_3|^2=1$.

We express the two qutrits time evolved flavour neutrino state as $\ket{\nu_{AB}(\theta,\phi,\eta,\xi_1,\xi_2)}=\ket{\nu_A(\theta,\phi,\eta,\xi_1,\xi_2)}\otimes\ket{\nu_B(\theta,\phi,\eta,\xi_1,\xi_2))}$, and find its the density matrix in the two qutrit standard basis  $\{\ket{11},\ket{12},\ket{13},\ket{21},\ket{22},\ket{23},\ket{31},\ket{32},\ket{33}\}\in\mathcal{H}^3_A\otimes\mathcal{H}^3_B$ as
\begin{eqnarray}
{\rho^{AB}_{9\times9}=\rho^A_{3\times3}\otimes \rho^B_{3\times3}=\ket{\nu_{AB}(\theta,\phi,\eta,\xi_1,\xi_2)}\bra{\nu_{AB}(\theta,\phi,\eta,\xi_1,\xi_2)}}&\nonumber\\
{=\begin{pmatrix}
      |\alpha_1|^2 & \alpha_1\alpha_2^* & \alpha_1\alpha^*_3 \\
      \alpha_2\alpha_1^* &  |\alpha_2|^2 & \alpha_2\alpha^*_3 \\
      \alpha_3\alpha_1^* & \alpha_3\alpha^*_2 & |\alpha_3|^2 \\
      \end{pmatrix}\otimes \begin{pmatrix}
      |\alpha^\prime_1|^2 & \alpha^\prime_1{\alpha^\prime_2}^* & {\alpha^\prime_1}{\alpha^\prime}^*_3 \\
      \alpha^\prime_2{\alpha^\prime}_1^* &  |\alpha^\prime_2|^2 & \alpha^\prime_2{\alpha^\prime}^*_3 \\
      \alpha^\prime_3{\alpha^\prime}_1^* & \alpha^\prime_3{\alpha^\prime}^*_2 & |\alpha^\prime_3|^2 \\
      \end{pmatrix}=( ... )_{9X9}}&
      \label{81}
\end{eqnarray}
 where $\rho^A_{3\times3}=$ $\ket{\nu_A(\theta,\phi,\eta,\xi_1,\xi_2)}$  $\bra{\nu_A(\theta,\phi,\eta,\xi_1,\xi_2)}$ and \\
  $\rho^B_{3\times3}=\ket{\nu_B(\theta,\phi,\eta,\xi_1,\xi_2)}$ $\bra{\nu_B(\theta,\phi,\eta,\xi_1,\xi_2)}$ are the density matrix of two sub-systems A and B, respectively. Also, $\alpha^*_1,\alpha^*_2,\alpha^*_3$ and ${\alpha^\prime}^*_1,{\alpha^\prime}^*_2,{\alpha^\prime}^*_3$ are complex conjugate of $\alpha_1,\alpha_2,\alpha_3$ and $\alpha^\prime_1,\alpha^\prime_2,\alpha^\prime_3$, respectively.

 Alternatively, the density matrix in Eq.(\ref{81}) can be expanded uniquely as 
\begin{eqnarray}
{\rho^{AB}_{9\times9}=\frac{1}{3}(I+\sqrt{3}\hat{n}.\Vec{\lambda}^A)\otimes\frac{1}{3}(I+\sqrt{3}\hat{n}^\prime.\Vec{\lambda}^B)}&\nonumber\\
{=\frac{1}{9}(I\otimes I+\sqrt{3}\vec{\lambda}^A.\hat{n}\otimes I+\sqrt{3} I \otimes \vec{\lambda}^{B}.\hat{n}^\prime + \frac{3}{2} \sum_{i,j=1}^8c_{ij}\lambda^A_i\otimes\lambda^{B}_j).\hspace{0.7cm}}&
\label{Eq.74}
\end{eqnarray}

The (real) expansion coefficients in Eq.(\ref{Eq.74}) are given by

\begin{eqnarray}
{n_i=\frac{\sqrt{3}}{2} tr(\rho^{AB}_{9\times 9}\lambda_i\otimes I),}&\nonumber\\
{n^\prime_j=\frac{\sqrt{3}}{2} tr(\rho^{AB}_{9\times 9} I\otimes \lambda_j),}&\nonumber\\
{c_{ij}=\frac{3}{2} tr(\rho^{AB}_{9\times 9} \lambda_i\otimes\lambda_j)},
\end{eqnarray}
where $n_i$ and $n_j$ are components of unit vector $\hat{n}$ and $\hat{n}^\prime$ of the two sub-systems: $\rho^A_{3\times3}$ and $\rho^{B}_{3\times3}$ and $i,j=1,...,8$. The coefficients $c_{ij}$ form a $8\times8$ correlation matrix $R$. 
The two qutrit density matrix shown in Eq.(\ref{Eq.74}) can be also cast into the form as
\begin{eqnarray}
\rho^{AB}_{9\times9}=\rho^A_{3\times3}\otimes{\rho}^{B}_{3\times3}=(\frac{1}{3}c_\alpha\lambda_\alpha)\otimes (\frac{1}{3}c_{\beta}\lambda_{\beta})=\frac{1}{9}c_{\alpha\beta} \lambda_\alpha\otimes\lambda_{\beta},\nonumber\\
\end{eqnarray}
where the expansion coefficients are given by 
\begin{equation}
c_{\alpha\beta}=\frac{9}{4}tr(\rho^{AB}_{9\times9}\lambda_\alpha\otimes\lambda_\beta).
\label{Eq.77}
\end{equation}
$\alpha,\beta=0,...,8$ and normalization requires that $c_{00}=\frac{3}{2}$. Thus, Eq.(\ref{Eq.77}) form a generalized matrix $\textbf{GM}$ which is split into four components: a scalar of $\frac{3}{2}$, two eight-dimensional vectors, and a $8\times8$ correlation matrix R. We write
 \begin{equation}
\textbf{GM}=\left[\begin{array}{c|ccc}

\frac{3}{2} & c_{01} &.....& c_{08} \\

\hline

c_{10}&c_{11}&.....& c_{18}\\
: &:&.....&:\\
: &:&.....&:\\
c_{80}&c_{81}&.....& c_{88}\\
\end{array}\right ],
\label{86}
\end{equation}

 where $n_i=c_{i0}$ and $n^\prime_j=c_{0j}$ ($i,j=1,...,8$) are the components of local unit Bloch vectors $\hat{n}$ and $\hat{n}^{\prime}$, respectively of the two single qutrit sub-systems ($\rho^A_{3\times3}$ and $\rho^B_{3\times3}$), and $c_{ij}$ $(i,j=1,...,8)$ are the matrix elements of correlation matrix R. Therefore, according to the above Eq.(\ref{86}), all possible combination of neutrinos two qutrits density matrix  like $\rho^{ee}_{9\times9}=\rho^{e}_{3\times3}\otimes\rho^e_{3\times3}$, $\rho^{e\mu}_{9\times9}=\rho^{e}_{3\times3}\otimes\rho^\mu_{3\times3}$, $\rho^{\mu\mu}_{9\times9}=\rho^{\mu}_{3\times3}\otimes\rho^\mu_{3\times3}$, $\rho^{\mu e}_{9\times 9}=\rho^{\mu}_{3\times3}\otimes\rho^e_{3\times3}$, etc., resembles a separable state.
    
In general, the entanglement measure generalized concurrence for the two qutrits mixed state density matrix $\rho_{9\times9}$ is defined as \cite{Luthra}
\begin{equation}
C_3(\rho_{9\times9})=max\{0,2\mu_1-\sum^9_{i=1}\mu_i\},
\label{Eq.78}
\end{equation}
where the $\mu_i$ (with i=1,2,...,9) are the square roots of the eigenvalues of the non-Hermitian matrix $\rho_{9\times9}\tilde{\rho}_{9\times9}$ in decreasing order. $\tilde{\rho}_{9\times9}$ is the spin-flip density matrix
\begin{equation}
\tilde{\rho}_{9\times9}=(O_3 \otimes O_3){\rho^*}_{9\times9}(O_3\otimes O_3),
\label{Eq.79}
\end{equation}
 with ${\rho^*}_{9\times9}$ being the complex conjugate of $\rho_{9\times9}$ and $O_3$ is the transformation matrix for qutrits
 \begin{equation}
 O_3=\begin{pmatrix}
     0&-i&i\\
     i&0&-i\\
     -i&i&0\\
     \end{pmatrix}.
 \end{equation}
Here, $O_3\otimes O_3$ is analogous to the $\sigma_y\otimes\sigma_y$ in the two-qubit system (see Eq.(\ref{Eq.27})).
We find that the generalized concurrence for possible combination of two qutrits separable state density matrix is zero, i.e. $C_{3}(\rho^{ee}_{9\times9})=C_3(\rho^{e\mu}_{9\times9})=C_3(\rho^{\mu\mu}_{9\times9})=C_3(\rho^{\mu e}_{9\times 9})=...=0$.
   
Furthermore, to investigate two qutrits entanglement in neutrino oscillations, we study two-flavour neutrino oscillations in the bipartite qutrit system and quantify the generalized concurrence. We map the neutrino flavour state at t=0 to bipartite qutrit states as $\ket{\nu_e}=\ket{1}\otimes\ket{2}$ and $\ket{\nu_\mu}=\ket{2}\otimes\ket{1}$. Then using Eq.(\ref{37}), the normalized time evolved electron flavour neutrino state in linear superposition of bipartite qutrit neutrino flavour basis is
\begin{equation}
\ket{\nu_e (\theta,\phi)}_f=({\cos}^2\theta +{\sin}^2\theta e^{-i\phi})\ket{12}+{\sin}\theta {\cos}\theta (1-e^{-i\phi })\ket{21}\hspace{1cm}
\end{equation} 
and its density matrix ${\rho^e_{f}}_{9\times9}=\ket{\nu_e (\theta,\phi)}_f{}_f\bra{\nu_e (\theta,\phi)}$ in the two qutrit standard basis $\{\ket{11},\ket{12},\ket{13},\ket{21},$ $\ket{22},\ket{23},\ket{31},\ket{32},\ket{33}\}$ is

\begin{equation}
{\rho^e_{f}}_{9\times9}=\begin{pmatrix}
0&0&0&0&0&0&0&0&0\\
0&|a_1|^2&0&a_1b_1^{*}&0&0&0&0&0\\
0&0&0&0&0&0&0&0&0\\
0&a_1^{*}b_1&0&|b_1|^2&0&0&0&0&0\\
0&0&0&0&0&0&0&0&0\\
0&0&0&0&0&0&0&0&0\\
0&0&0&0&0&0&0&0&0\\
0&0&0&0&0&0&0&0&0\\
0&0&0&0&0&0&0&0&0\\
\end{pmatrix}
\label{90}
\end{equation}

where $a_1=({\cos}^2\theta +{\sin}^2\theta e^{-i\phi})$, and $b_1={\sin}\theta {\cos}\theta (1-e^{-i\phi })$ are the coefficients of bipartite qutrit flavor basis $\ket{\nu_e}$ and $\ket{\nu_\mu}$, respectively, and  $a_1^{*}$, $b_2^{*}$ are complex conjugate of $a_1$ and $b_1$, respectively.  

 \begin{figure}[H]
\begin{center}
\resizebox{1.0\columnwidth}{!}{
    \includegraphics{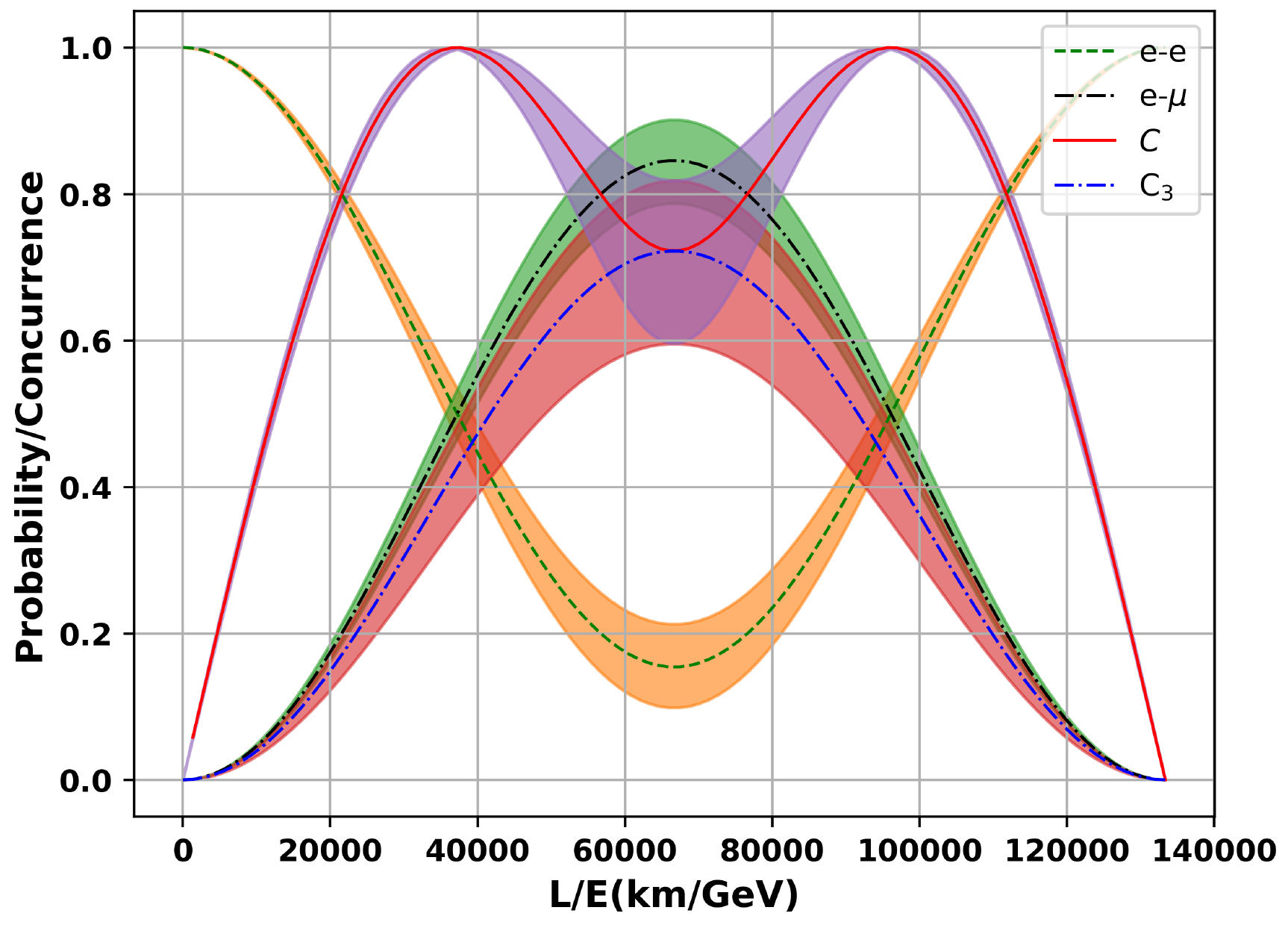} } 
  \caption{The violet band shows the $\nu_e$ concurrence $C{(\rho^e_f}_{4\times4})$ (Red, solid line) in the bipartite qubit system and the pink band shows the generalized concurrence $C_3({\rho^e_f}_{9\times9})$ (Blue, dash dotted line) in the bipartite qutrit system. Both entanglement measures are compared with the orange band which shows the $\nu_e$ probability $P_{e\rightarrow e}$ (Green, dashed line) and with the green band which shows the $P_{e\rightarrow \mu}$ probability (Black, dash dotted line), using the current experimental bounds of the $3\sigma$ range of neutrino parameters \cite{Esteban:2020cvm}.}
\label{fig:3}
\end{center}
\end{figure} 

Using Eq.(\ref{90}) in Eq.(\ref{Eq.79}), we construct the spin-flip density matrix ${\tilde{\rho^e_f}_{9\times9}}=(O_3 \otimes O_3){\rho^*_f}^e_{9\times9}(O_3\otimes O_3)$. We find that only one of the square root of eigenvalues of the matrix ${\rho^e_{f}}_{9\times9}{\tilde{\rho^e_f}_{9\times9}}$ is non-zero, i.e., $\mu_1=4{\cos}\theta {\sin}\theta ({\cos}^2\theta-{\sin}^2\theta){\sin}^2{\frac{\phi}{2}}$. Thus using Eq.(\ref{Eq.78}), the generalized concurrence of the time evolved electron flavour neutrino state $\ket{\nu_e (\theta,\phi)}_f$ in the bipartite qutrit system is quantified as 
\begin{equation}
C_3({\rho^e_f}_{9\times9})=4{\cos}\theta {\sin}\theta ({\cos}^2\theta-{\sin}^2\theta){\sin}^2{\frac{\phi}{2}}.
\label{91}
\end{equation}

   In Fig.(\ref{fig:3}), the generalized concurrence $C{(\rho^e_f}_{9\times9})$ (see Eq.(\ref{91})) of the time evolved electron flavour neutrino state in the bipartite qutrit system is compared with the concurrence $C{(\rho^e_f}_{4\times4})$ (see Eq.(\ref{40})) in the bipartite qubit system. Thus, the nonzero value of the generalized concurrence shows that in the two neutrino system, the time evolved neutrino flavour state is a bipartite qutrit entangled state. Therefore, the plot results warrant a study of two qutrits entanglement in the three-flavour neutrino oscillation.

\section{Discussion and conclusions} 
\label{Sec.6}
                                                                                                                                                                                                                                                                                                                                                                                                                                                                                                                                                                                                                                                                                                                                                                                                                                                                                                                                                                                                                                                                                                                                                                                                                                                                                                                                                                                                                                                                                                                                                                                                                                                                                                                                                                                                                                                                                                                                                                                                                                                                                                                                                                                                                                                                                                                                                                                          
   In this work, we use the Pauli matrices to characterize the two-flavour neutrino oscillations on the Poincar\'e sphere $S^2=SU(2)/U(1)$. It is shown that the Poincar\'e vector of the time evolved flavour neutrino state lies on the unit sphere in the three-dimensional real vector space. This result helps us to characterize the two neutrino system as qubits.
   
     In the two-qubit systems, we have shown the Poincar\'e sphere representation of two neutrino system. We constructed two-qubit density matrix of neutrinos in the basis of the Dirac matrices. The coefficients of the Dirac matrices form the Bloch matrix, which shows that the two-qubit neutrino state is a separable state. Furthermore, we map the mass eigenstates of neutrinos directly to the bipartite qubit system, where the Bloch matrix construction show that the bipartite qubit neutrino state is an entangled state. The quantification of entanglement measure concurrence in neutrino oscillation probabilities in the two neutrino system shows that time evolved flavour neutrino states are bipartite qubit entangled states.   
   
We use the Gell-Mann matrices to construct the Poincar\'e sphere $S^7=SU(3)/U(2)$ in the three-flavour neutrino oscillation. The SU(3) result allows us to identify the three neutrino system as qutrits which generalize the concept of entangled tripartite states of neutrinos. We calculate the entropy of mixing $E(\rho^e_d)$ of the time evolved flavour neutrino mixed state in a single qutrit system using the current experimental bound on the neutrino oscillation parameters, and we find that the equi-mixing curves of $E(\rho^e_d)$ lie inside the qutrit triangle.

The argument for using qutrits instead of qubits in quantum entanglement for computational purposes is emphasized by the demonstrated advantage of entangled qutrits being less affected by noise compared to entangled qubits. This is supported by research, such as the paper referenced \cite{Daniel: 2002}, which shows that bipartite systems of higher dimensionality, such as qutrits, exhibit a greater resistance to noise and a maximum violation of Bell's Inequality. Therefore, developing a theory of entanglement for neutrinos based on qutrits is not only justified but also potentially beneficial.

In the two-qutrit system, constructing a generalized Poincar\'e sphere using the Gell-Mann matrix tensor products led to the generalized Bloch matrix in the Bloch vector space of the three neutrino system. The quantification of the generalized concurrence in the two neutrino system implies that the two flavour neutrino oscillations are bipartite qutrit entangled states. We have compared the generalized concurrence of the bipartite qutrit neutrinos with the concurrence of the bipartite qubit neutrino. Both measures provide a qualitatively nonzero amount of information in the two neutrino system. In a subsequent study, we shall examine two qutrit entanglement in the three neutrino system \cite{Jafarizadeh:2008}.

  A quantum computer did the quantum simulation of bipartite qubit entanglement of two-flavour neutrino oscillations \cite{Jha:2021}. Recently, the simulation of the coherent collective oscillations of a system of N neutrinos in the two-flavour approximation was examined using the IBM quantum device based on trapped-ions qubits \cite{Amitrano:2022yyn}. However, new studies claim that qutrits offer a promising path towards extending the frontier of quantum computers \cite{Gokhale:2020, Li:2013}. The use of qutrits in studying N-neutrino interactions in the astrophysical scenario could be particularly useful \cite{Siwach:2022xhx}. In future, qutrit computers can be used to delve into quantum information studies of neutrinos and explore the intricacies of entanglement in collective three-flavour neutrino oscillations. This could provide valuable insights into the behaviour and properties of neutrinos in astrophysical environments. 
  
In our paper, we associated three-flavour neutrinos to qutrit system. So, by using quantum gates and circuit given in ref.\cite{Gokhale:2020}, one can simulate the neutrino flavor oscillations in qutrit system, which can be used to investigate the properties of neutrinos oscillations and their entanglement. Thus, the results of our paper lead us to a new direction of ternary computing using qutrits.

\section{Acknowledgments}
AKJ acknowledges a project funded by SERB, India, with Ref. No. CRG/2022/003460, for partial support towards this research. AC would like to acknowledge the support from DST for this work through the project : DST/02/0201/2019/01488.

\section{Data availability}
The manuscript has no associated data.

\end{document}